\documentclass{IEEEtran}
\usepackage{cite}
\usepackage{amsmath,amssymb,amsfonts}
\usepackage{algorithmic}
\usepackage{graphicx}
\usepackage{soul}
\usepackage{xcolor}
\usepackage{textcomp}
\usepackage{lipsum}

\ifCLASSOPTIONcompsoc
    \usepackage[caption=false, font=normalsize, labelfont=sf, textfont=sf]{subfig}
\else
    \usepackage[caption=false, font=footnotesize]{subfig}
\fi

\def\BibTeX{{\rm B\kern-.05em{\sc i\kern-.025em b}\kern-.08em
    T\kern-.1667em\lower.7ex\hbox{E}\kern-.125emX}}
\begin{document}
\title{Characterization of a Displaced Coaxial Feed for Cascaded Cylindrical Metasurfaces}

\author{Chun-Wen Lin, \IEEEmembership{Graduate Student Member, IEEE}, Richard W. Ziolkowski, \IEEEmembership{Life Fellow, IEEE}, and Anthony Grbic, \IEEEmembership{Fellow, IEEE}

\thanks{This work has been submitted to the IEEE for possible publication. Copyright may be transferred without notice, after which this version may no longer be accessible.}

\thanks{This work was supported by the Air Force Office of Scientific Research (AFOSR) Multidisciplinary University Research Initiative (MURI) program under Grant FA9550-18-1-0379, and the UM-KACST Joint Center for Microwave Sensor Technology \textit{(Corresponding Author: Anthony Grbic.)}}

\thanks{Chun-Wen Lin and Anthony Grbic are with the Radiation Laboratory, Department of Electrical Engineering and Computer Science, University of Michigan, Ann Arbor, MI 48109-2122 USA (email: chunwen@umich.edu; agrbic@umich.edu)}

\thanks{Richard W. Ziolkowski is with the Department of Electrical and Computer Engineering, University of Arizona, Tucson, AZ 85721-0104 USA (e-mail: ziolkows@arizona.edu)}
}

\maketitle


\begin{abstract}
This paper characterizes a realistic feed for cylindrical metasurfaces, allowing it to be included in metasurface design.
Specifically, it investigates a coaxial feed which is displaced from the center (off-center) of concentrically-cascaded cylindrical metasurfaces.
Formulas are reported to quickly compute the multimodal $S$-matrix (scattering properties) of a displaced feed from that of the central feed.
The theory is rigorously derived based on the addition theorem of Hankel functions for all azimuthal modes.
Moreover, the resulting multimodal $S$-matrix is combined with the multimodal wave matrix theory used to model cylindrical metasurfaces,
allowing devices to be designed that realize arbitrary field transformations from a displaced coaxial feed.
A design example is reported, which opens new opportunities in the realization of realistic, high-performance cylindrical-metasurface-based devices.
\end{abstract}



\section{Introduction}
\label{sec:introduction}
\IEEEPARstart{C}{oncentrically}-cascaded cylindrical metasurfaces have been demonstrated that can tailor the amplitude and  phase \cite{Xu_PRA_2020, Xu_APS_2020, my_TAP_2023, my_APS_2021}, and even the polarization \cite{Liu_TAP_2021, my_TAP_2021} of cylindrical waves.
As a result, they have been employed in stealth technology, providing functionalities such as camouflage \cite{Safari_PRB_2019}, electromagnetic cloaking \cite{Chen_PRB_2011, Selvanayagam_AWPL_2012, Selvanayagam_PRX_2013, Sounas_PRA_2015, Kwon_PRB_2018, Lee_PRA_2022, Dehmollaian_PRA_2023} and illusion \cite{Xu_PRA_2020, Xu_APS_2020, my_TAP_2023, Safari_PRB_2019, Kwon_PRB_2020}.
They have also been used to manipulate the wave propagation \cite{Mazor_MTM_2022} and cutoff frequency \cite{Barker_MTM_2020, Barker_TMTT_2023} in circular waveguides, and to design a high gain antenna by shaping the radiation patterns \cite{Xu_PRA_2020, my_EuCAP_2023}.
In addition, cylindrical metasurfaces can be used to generate orbital angular momentum (OAM) waves through azimuthal mode conversion \cite{my_TAP_2023, my_APS_2021, my_AWPL_2023, my_APS_2022, Li_PRA_2019, Mazor_PRB_2019}.
All the aforementioned applications require arbitrary field transformations which convert a known excitation field to the desired or stipulated output field.

Previous approaches to achieve arbitrary field transformations through cascaded cylindrical metasurfaces were mostly based on the Generalized Sheet Transition Condition (GSTC) \cite{Smy_TMTT_part1_2023, Smy_TMTT_part2_2023, Smy_APS_2022, Sandeep_JMMCT_2018, Sandeep_JEWA_2023, Sipus_RadioE_2019}, which suffers greatly from realization limitations such as extremely small layer separations \cite{Xu_PRA_2020, Smy_TMTT_part1_2023, Smy_TMTT_part2_2023} an the need for perfectly conducting baffles between unit cells \cite{Xu_PRA_2020, Li_PRA_2019}.
In contrast, multimodal wave matrix theory \cite{my_TAP_2023, my_APS_2021} is able to accurately capture the higher-azimuthal-order coupling and lateral wave propagation between the cascaded cylindrical metasurface layers, making it ideal for analyzing and synthesizing such structures.
Using the multimodal wave matrix theory, an azimuthal mode converter and an illusion device have been designed \cite{my_TAP_2023}.
However, in these earlier works, idealized (fictitious) line currents were assumed as the excitation sources.
This impractical assumption neglects possible interaction between the source and metasurfaces.

\begin{figure}[!t]
\centering
\subfloat[\footnotesize]{%
    \includegraphics[width=0.5\linewidth]{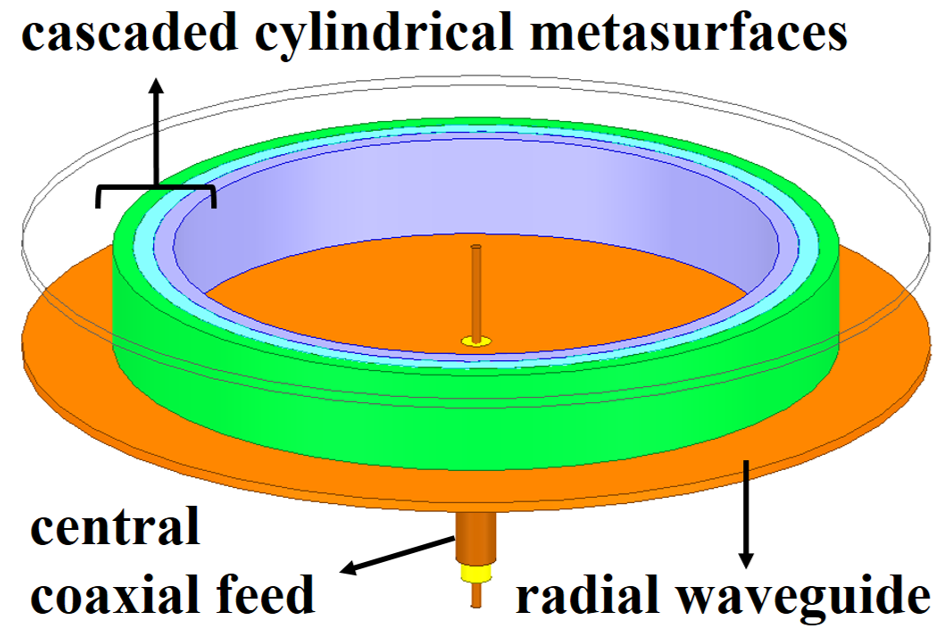}}
\subfloat[\footnotesize]{%
    \includegraphics[width=0.5\linewidth]{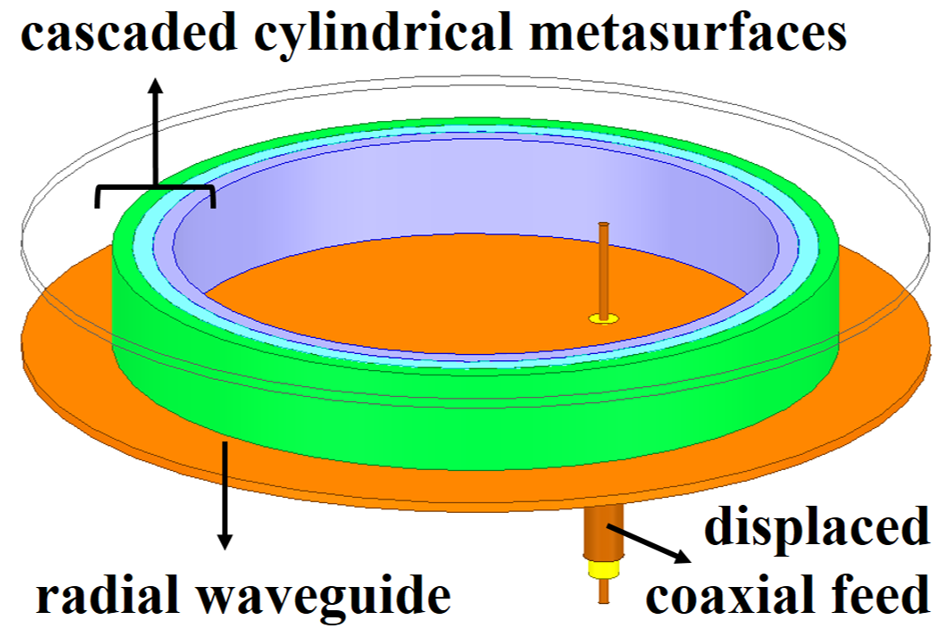}}
\caption{A practical cylindrical-metasurface-based device excited by a realistic coaxial feed. For clarity, the upper plate of radial waveguide is made transparent. Illustrations of the cases where the coaxial feed is (a) located at the center of the metasurfaces, and (b) displaced from the center.}
\label{fig_intro}
\end{figure}

Subsequently, a realistic device, shown in Fig. \ref{fig_intro}(a), has been developed, which consists of a set of cascaded cylindrical metasurfaces inside a radial waveguide with a realistic, central coaxial feed \cite{my_AWPL_2023, my_APS_2022}.
The scattering properties of the realistic, central coaxial feed are numerically computed using the mode-matching technique (MMT) \cite{Shen_MOTL_1999, Heebl_PRA_2016, Faris_OJAP_2021, Eleftheriades_TMTT_1994, Kuhn_AEU_1973} and compactly summarized into a multimodal $S$-matrix.
This multimodal $S$-matrix is then integrated with the multimodal wave matrix representing the metasurfaces \cite{my_TAP_2023, my_APS_2021}.
This allows interactions between the cylindrical metasurfaces and the central coaxial feed to be taken into account.
Based on these structures, a coaxially-fed, azimuthal mode converter was realized \cite{my_AWPL_2023, my_APS_2022}.
These single-input single-output (SISO) devices are able to perform arbitrary field transformations from a central coaxial feed.

Multiple-input multiple-output (MIMO) devices made from cylindrical metasurfaces can provide for greater functionalities.
MIMO cylindrical metasurfaces can be designed to interact with feeds at different locations and produce different output fields.
In other words, these MIMO devices incorporate multiple feeds, while only a single, centrally-located coaxial feed for cylindrical metasurfaces was investigated in \cite{my_AWPL_2023, my_APS_2022}.
Therefore, as a first step toward designing MIMO devices, a realistic coaxial feed displaced from the center of a cylindrical metasurface, illustrated in Fig. \ref{fig_intro}(b), must be modelled.
In \cite{my_EuCAP_2023}, the scattering characteristics of a displaced coaxial feed were computed.
A beam-shaping shell, consisting of five azimuthally-symmetric cylindrical metasurfaces, was presented, demonstrating the manipulation of electromagnetic fields from a displaced coaxial feed.

In this paper, the theory behind \cite{my_EuCAP_2023} is detailed.
To model a displaced coaxial feed, our aim is to derive its multimodal $S$-matrix since this mathematical form is compatible with the multimodal wave matrix theory used to model concentric metasurfaces \cite{my_TAP_2023, my_APS_2021}.
Previously, MMT was adopted in order to accurately model the central coaxial feed \cite{my_AWPL_2023}.
However, this method becomes computationally expensive when the feed is off-center.
Herein, we present a more elegant and efficient way to characterize the scattering properties 
of a displaced coaxial feed.
First, the multimodal $S$-matrices of the central feed and a displaced feed are defined.
Next, the addition theorems of Hankel functions, including all higher-order azimuthal modes \cite{Chew_book, Martin_book}, are then employed to relate these two matrices.
We have shown that the multimodal $S$-matrix of the displaced feed can be derived directly from the known matrix of the central feed in \cite{my_AWPL_2023, my_APS_2022}.
They are related through simple matrix multiplication.
With this result, cascaded cylindrical metasurfaces can be designed to generate any stipulated output field with flexibility over the feed placement.
To verify the proposed theory, a beam shaping shell fed by a displaced coaxial feed is presented in this paper.
The reflection coefficient of the device is reduced by introducing a quarter-wave impedance transformer to the coaxial feed \cite{Pozar_book, Ettorre_TAP_2012}.
This example has two significant implications. First, it showcases that arbitrary field transformations from a displaced coaxial feed can be accomplished.
More importantly, it creates a path toward the realization of practical cylindrical-metasurface-based MIMO devices. 


\section{The Proposed Theory}
\label{sec:theory}
The analysis and synthesis of cylindrical metasurface-based devices excited by a displaced coaxial feed
are presented in this section. Derivations of the associated multimodal {\it S}-matrices and consequent field representations are given.
The characterization of the displaced coaxial feed is discussed in detail.

\subsection{Proposed Structure and Definitions}

\begin{figure}[!t]
\centering
\subfloat[\footnotesize]{%
    \includegraphics[width=0.8\linewidth]{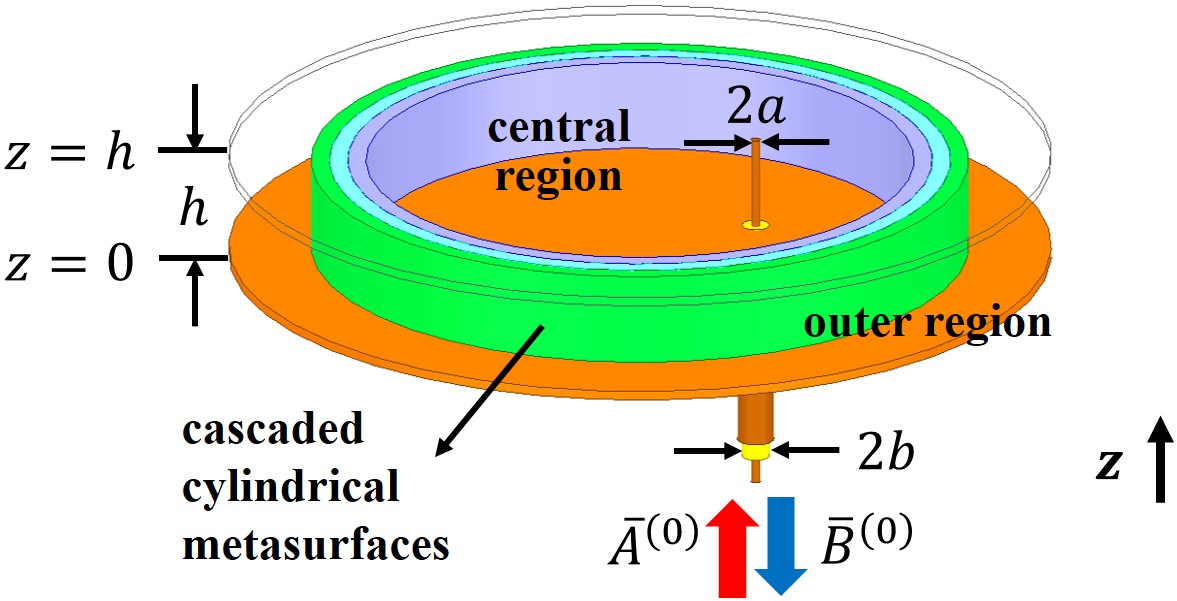}}
\\
\subfloat[\footnotesize]{%
    \includegraphics[width=0.8\linewidth]{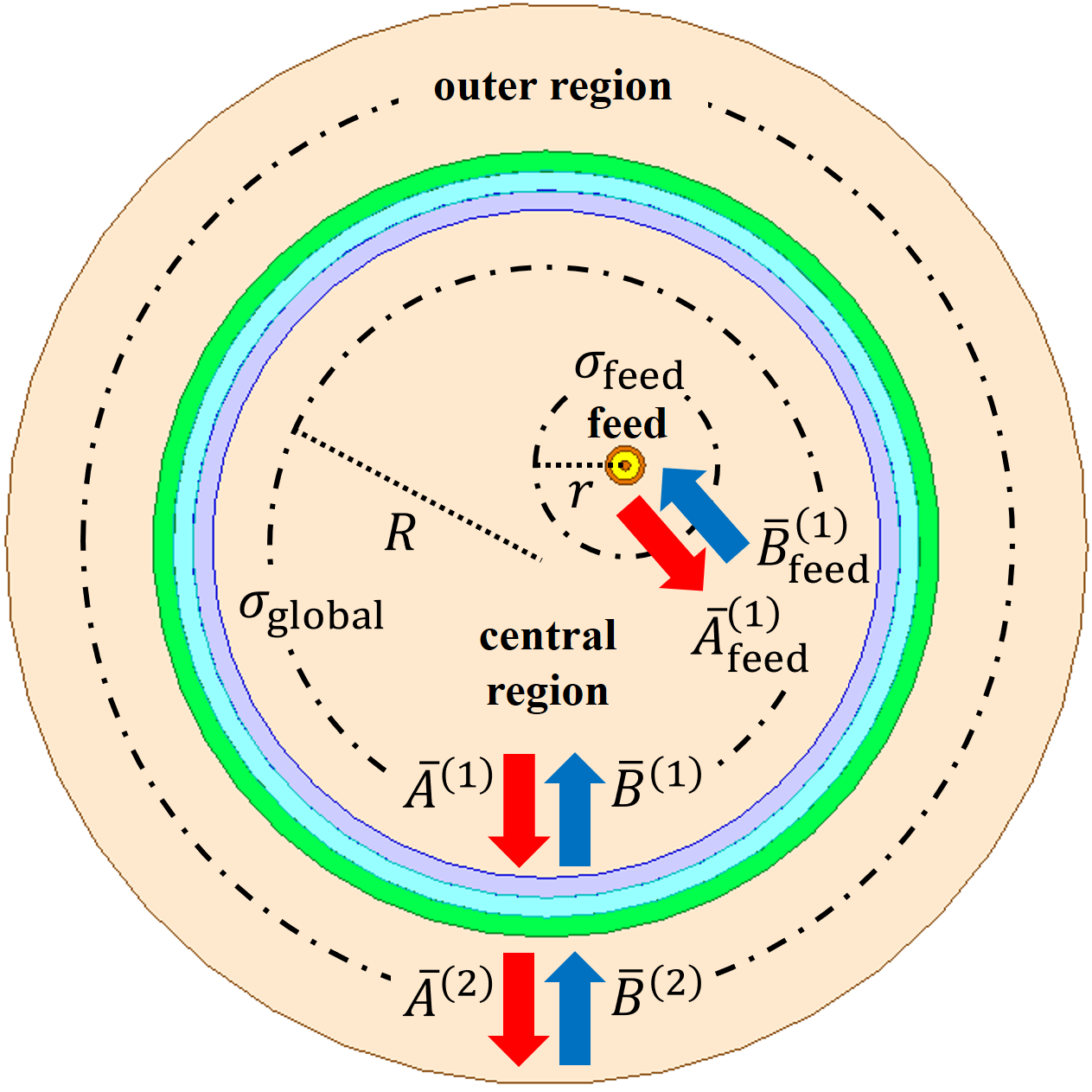}}
\caption{The developed cylindrical structure with an off-center coaxial feed. (a) Perspective view. (b) Top view. The cylindrical surface $\sigma_\text{feed}$ is centered about the displaced feed location $(\rho', \phi')$ with radius $r$. The cylindrical surface $\sigma_\text{global}$ is centered about the global origin with radius $R$.}
\label{fig_S_def}
\end{figure}

The device considered is illustrated in Fig. \ref{fig_S_def}(a) and details of its layout are given in
Fig. \ref{fig_S_def}(b).
A set of concentrically-cascaded cylindrical metasurfaces, whose center is set to be the global origin, is inserted within an air-filled parallel-plate radial waveguide.
For simplicity, the height of the radial waveguide, $h$, is subject to the following limitation:
\begin{equation}
    h<\frac{\pi}{\omega\sqrt{\mu_0\varepsilon_0}},
\label{eq_2A_h_limitation}
\end{equation}
\noindent{where $\omega=2\pi f$ is the angular frequency under the operating frequency, and $\varepsilon_0$ and $\mu_0$ are the free-space permittivity and permeability, respectively.
This limitation (\ref{eq_2A_h_limitation}) ensures that all the propagating fields in the radial waveguide must be $\text{TM}_z$ modes and invariant in $z$ \cite{my_AWPL_2023}.
Consequently, the electromagnetic problem becomes two-dimentional (2-D) \cite{my_TAP_2023}.
The device is excited by a realistic coaxial feed which is located off-center at $(\rho', \phi')$.
We assume in the same manner as in \cite{my_AWPL_2023} that the inner conductor of the feed touches
the upper plate of the radial waveguide.
To guarantee that only the $\text{TEM}_z$ mode propagates in the coaxial cable, its inner and outer conductor radii $a$, $b$, as well as the filling dielectric permittivity $\varepsilon$, must satisfy \cite{Pozar_book}:}
\begin{equation}
    a+b<\frac{2}{\omega\sqrt{\mu_0\varepsilon}}.
\label{eq_2A_ab_limitation}
\end{equation}

In order to characterize the junction between the coaxial feed and the radial waveguide, a multimodal $S$-matrix can be defined.
To this end, the electric field along the cylindrical surface $\sigma_\text{feed}$ indicated in
Fig. \ref{fig_S_def}(b), which is centered about the feed location $(\rho', \phi')$, is investigated first.
The radius $r$ of $\sigma_\text{feed}$ is chosen such that evanescent waves generated from the junction discontinuity have decayed significantly.
Under the $e^{+j\omega t}$ time convention, the electric field along $\sigma_\text{feed}$ can be written in terms of propagating cylindrical waves (Hankel functions) centered with respect to $(\rho', \phi')$:
\begin{equation}
\begin{split}
    E_z\Big{|}_{\text{along }\sigma_\text{feed}}
    &= \sum_{m=-M}^{+M} \alpha_{\text{feed},m}^+ H_m^{(2)}(k_0 r) e^{-jm\phi''} \\
    &+ \sum_{m=-M}^{+M} \alpha_{\text{feed},m}^- H_m^{(1)}(k_0 r) e^{-jm\phi''}
\end{split}
\label{eq_2A_Ez_sigma_feed}
\end{equation}
\noindent{where $k_0=\omega\sqrt{\mu_0\varepsilon_0}$ is the wave number in the free-space, $\phi''$ is the azimuthal angle with respect to $(\rho', \phi')$, and $M$ is a sufficiently large number that ensures the series (\ref{eq_2A_Ez_sigma_feed}) converges.}
The first and second series in (\ref{eq_2A_Ez_sigma_feed}) represent the outgoing and incoming fields, respectively, and $\alpha_{\text{feed},m}^\pm$ are their corresponding wave amplitudes \cite{my_TAP_2023}.

A multimodal wave matrix can be defined based on vectors 
representing the complex amplitudes of the azimuthal modes of (\ref{eq_2A_Ez_sigma_feed}) \cite{my_TAP_2023}.
However, a multimodal $S$-matrix is defined based on power waves \cite{my_TAP_2023, my_AWPL_2023};
it requires the normalization of the azimuthal modes \cite{Pozar_book, Kurokawa_TMTT_1965}.
Specifically, we can write the following power waves:
\begin{align}
    \bar{A}^{(1)}_\text{feed} &= \bar{\bar{c}}_r^+ \bar{\bar{H}}_r^+ \bar{\alpha}_\text{feed}^+
    \label{eq_2A_A1feed_def} \\
    \bar{B}^{(1)}_\text{feed} &= \bar{\bar{c}}_r^- \bar{\bar{H}}_r^- \bar{\alpha}_\text{feed}^-
    ,\label{eq_2A_B1feed_def}
\end{align}
\noindent{where the $N\times 1$ vectors ($N=2M+1$)}
\begin{align}
    \bar{\alpha}_\text{feed}^+ &=[\alpha_{\text{feed},+M}^+, \dots, \alpha_{\text{feed},0}^+, \dots, \alpha_{\text{feed},-M}^+]^T
    \label{eq_2A_alpha_feed_p} \\
    \bar{\alpha}_\text{feed}^- &=[\alpha_{\text{feed},+M}^-, \dots, \alpha_{\text{feed},0}^-, \dots, \alpha_{\text{feed},-M}^-]^T
    \label{eq_2A_alpha_feed_n}
\end{align}
\noindent{contain the wave amplitude of each azimuthal mode.
The $N\times N$ diagonal matrices $\bar{\bar{H}}_r^+$ and $\bar{\bar{H}}_r^-$ consist of Hankel functions:}
\begin{align}
    \bar{\bar{H}}_r^+ &= \text{diag}
    \Big{(} H^{(2)}_{+M}(k_0 r),\dots,H^{(2)}_{0}(k_0 r),\dots,H^{(2)}_{-M}(k_0 r) \Big{)}
    \label{eq_2A_H_feed_p} \\
    \bar{\bar{H}}_r^- &= \text{diag}
    \Big{(} H^{(1)}_{+M}(k_0 r),\dots,H^{(1)}_{0}(k_0 r),\dots,H^{(1)}_{-M}(k_0 r) \Big{)},
    \label{eq_2A_H_feed_n}
\end{align}
\noindent{and the $N\times N$ diagonal matrices $\bar{\bar{c}}_r^+$ and $\bar{\bar{c}}_r^-$ are formed by
the normalization constants \cite{my_TAP_2023}:}
\begin{align}
    \bar{\bar{c}}_r^+ &= \text{diag}
    \Big{(} c_{r, +M}^+,\dots,c_{r, 0}^+,\dots,c_{r, -M}^+ \Big{)}
    \label{eq_2A_c_feed_p} \\
    \bar{\bar{c}}_r^- &= \text{diag}
    \Big{(} c_{r, +M}^-,\dots,c_{r, 0}^-,\dots,c_{r, -M}^- \Big{)}
    \label{eq_2A_c_feed_n}
\end{align}
\begin{align}
    c_{r, m}^+ &= \sqrt{2\pi rh \> \text{Re}
    \Big{\{}
    \frac{j\omega\varepsilon_0}{k_0}
    \frac{H_m^{(2)'}(k_0 r)}{H_m^{(2)}(k_0 r)}
    \Big{\}}}
    \label{eq_2A_c_feed_p_ind} \\
    c_{r, m}^- &= \sqrt{2\pi rh \> \text{Re}
    \Big{\{}
    \frac{j\omega\varepsilon_0}{k_0}
    \frac{H_m^{(1)'}(k_0 r)}{H_m^{(1)}(k_0 r)}
    \Big{\}}}
    \label{eq_2A_c_feed_n_ind}.
\end{align}
\noindent{The normalization constants (\ref{eq_2A_c_feed_p_ind}) and (\ref{eq_2A_c_feed_n_ind}) of an azimuthal mode are found by examining the powers that pass through a cylindrical reference surface \cite{my_TAP_2021}.
These constants take into account the fact that wave impedances in cylindrical coordinates depend on both the azimuthal order $m$ and the radius $r$.
Note that the power waves $\bar{A}^{(1)}_\text{feed}$ and $\bar{B}^{(1)}_\text{feed}$ are
$N\times 1$ vectors. They are defined on $\sigma_\text{feed}$ and, hence, are centered
with respect to $(\rho',\phi')$, as illustrated in Fig. \ref{fig_S_def}(b).}

The normalized waves in the coaxial cable that are incident on and reflected from the junction are denoted by $\bar{A}^{(0)}$ and $\bar{B}^{(0)}$, respectively (see Fig. \ref{fig_S_def}(b)). They are $1\times 1$ vectors since only the $\text{TEM}_z$ mode is allowed. Normalization of these two vectors is discussed in \cite{my_AWPL_2023, Pozar_book}.
For convenience, the reference surface of the power waves $\bar{A}^{(0)}$ and $\bar{B}^{(0)}$
is defined at the lower plate of the coaxial waveguide, i.e.
at the $z=0$ plane indicated in Fig. \ref{fig_S_def}(a) \cite{my_AWPL_2023}.
The multimodal $S$-matrix of the coaxial-waveguide junction can then be defined as,
\begin{equation}
    \begin{bmatrix}
    \bar{B}^{(0)} \\ \bar{A}^{(1)}_\text{feed}
    \end{bmatrix}
    =
    \begin{bmatrix}
    \bar{\bar{S}}_{f,11} & \bar{\bar{S}}_{f,12} \\ \bar{\bar{S}}_{f,21} & \bar{\bar{S}}_{f,22}
    \end{bmatrix}
    \cdot
    \begin{bmatrix}
    \bar{A}^{(0)} \\ \bar{B}^{(1)}_\text{feed}
    \end{bmatrix}.
\label{eq_2A_Sf_def}
\end{equation}
\noindent{Consequently, these power waves (\ref{eq_2A_Sf_def}) are centered
with respect to the coaxial feed.
Therefore, this definition (\ref{eq_2A_Sf_def}) is exactly the multimodal $S$-matrix of the central feed derived
in \cite{my_AWPL_2023} through MMT.}

Similarly, we can define a multimodal $S$-matrix of the displaced feed,
\begin{equation}
    \begin{bmatrix}
    \bar{B}^{(0)} \\ \bar{A}^{(1)}
    \end{bmatrix}
    =
    \begin{bmatrix}
    \bar{\bar{S}}_{d,11} & \bar{\bar{S}}_{d,12} \\ \bar{\bar{S}}_{d,21} & \bar{\bar{S}}_{d,22}
    \end{bmatrix}
    \cdot
    \begin{bmatrix}
    \bar{A}^{(0)} \\ \bar{B}^{(1)}
    \end{bmatrix},
\label{eq_2A_Sd_def}
\end{equation}
\noindent where $\bar{A}^{(1)}$ and $\bar{B}^{(1)}$ are both $N\times 1$ vectors defined on the
cylindrical surface $\sigma_\text{global}$ shown in Fig. \ref{fig_S_def}(b).
The radius $R$ of this cylindrical surface is chosen such that $\sigma_\text{global}$ fully encloses $\sigma_\text{feed}$, namely, $R>\rho'+r$.
The power waves $\bar{A}^{(1)}$ and $\bar{B}^{(1)}$ are centered with respect to the global origin.
These waves are the natural bases of the cylindrical metasurfaces, so they greatly simplify any further analysis.

Analogous to (\ref{eq_2A_A1feed_def}) and (\ref{eq_2A_B1feed_def}), we write:
\begin{align}
    \bar{A}^{(1)} &= \bar{\bar{c}}_R^+ \bar{\bar{H}}_R^+ \bar{\alpha}_\text{global}^+
    \label{eq_2A_A1_def} \\
    \bar{B}^{(1)} &= \bar{\bar{c}}_R^- \bar{\bar{H}}_R^- \bar{\alpha}_\text{global}^-
    ,\label{eq_2A_B1_def}
\end{align}
\noindent in which the $N\times N$ diagonal matrices $\bar{\bar{c}}_R^\pm$ and $\bar{\bar{H}}_R^\pm$ are similarly defined as in (\ref{eq_2A_H_feed_p})-(\ref{eq_2A_c_feed_n_ind}), with the argument $r$ replaced with $R$.
The $N\times 1$ vectors $\bar{\alpha}_{\text{global}}^\pm$ are composed of wave amplitudes obtained by expanding the electric field along $\sigma_\text{global}$:
\begin{equation}
\begin{split}
    E_z\Big{|}_{\text{along }\sigma_\text{global}}
    &= \sum_{m=-M}^{+M} \alpha_{\text{global},m}^+ H_m^{(2)}(k_0 R) e^{-jm\phi} \\
    &+ \sum_{m=-M}^{+M} \alpha_{\text{global},m}^- H_m^{(1)}(k_0 R) e^{-jm\phi}
\end{split}
\label{eq_2A_Ez_sigma_global}
\end{equation}
\begin{align}
    \bar{\alpha}_\text{global}^+ &=[\alpha_{\text{global},+M}^+, \dots, \alpha_{\text{global},0}^+, \dots, \alpha_{\text{global},-M}^+]^T
    \label{eq_2A_alpha_global_p} \\
    \bar{\alpha}_\text{global}^- &=[\alpha_{\text{global},+M}^-, \dots, \alpha_{\text{global},0}^-, \dots, \alpha_{\text{global},-M}^-]^T
    \label{eq_2A_alpha_global_n}.
\end{align}
where $\phi$ is the azimuthal angle measured from the global origin.
Furthermore, the $N\times 1$ vectors $\bar{A}^{(2)}$ and $\bar{B}^{(2)}$ are defined in the outer region
of Fig. \ref{fig_S_def}(b). They are related to $\bar{A}^{(1)}$ and $\bar{B}^{(1)}$ through the multimodal $S$-matrix of the cascaded cylindrical metasurfaces:
\begin{equation}
    \begin{bmatrix}
    \bar{B}^{(1)} \\ \bar{A}^{(2)}
    \end{bmatrix}
    =
    \begin{bmatrix}
    \bar{\bar{S}}_{M,11} & \bar{\bar{S}}_{M,12} \\ \bar{\bar{S}}_{M,21} & \bar{\bar{S}}_{M,22}
    \end{bmatrix}
    \cdot
    \begin{bmatrix}
    \bar{A}^{(1)} \\ \bar{B}^{(2)}
    \end{bmatrix}.
\label{eq_2A_Sm_def}
\end{equation}
\noindent For a given cascaded cylindrical metasurface, its multimodal $S$-matrix can be derived based on the multimodal wave matrix theory detailed in \cite{my_TAP_2023}. 

\subsection{Analysis and Synthesis}

\begin{figure}[!t]
\centering
\subfloat[\footnotesize]{%
    \includegraphics[width=7.6cm]{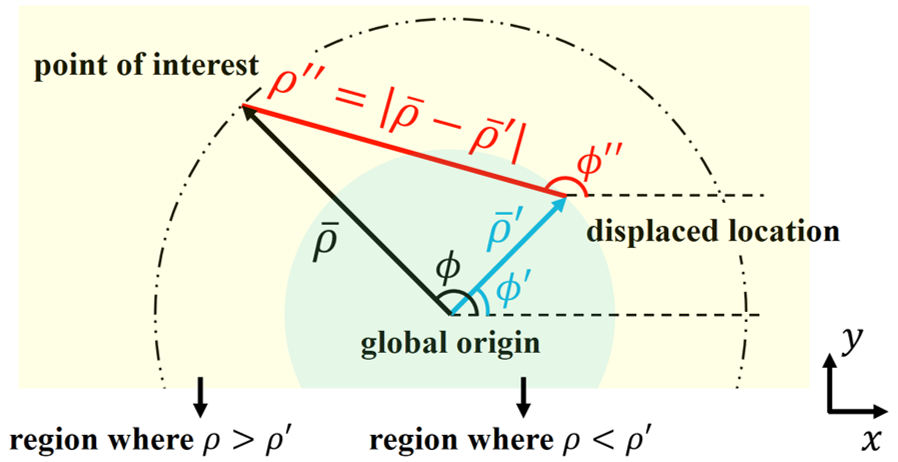}}
\\
\subfloat[\footnotesize]{%
    \includegraphics[width=7.6cm]{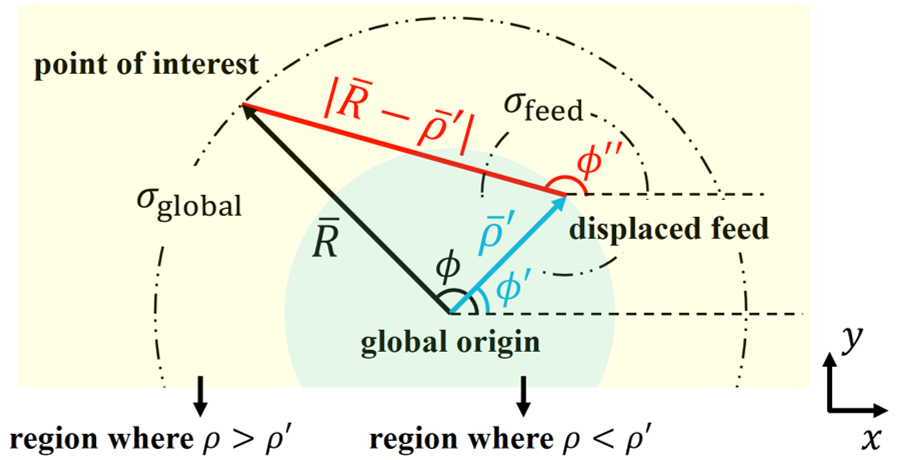}}
\caption{Addition theorem details.
(a) Geometric parameters in the addition theorem for Hankel functions (\ref{eq_2B_Hankel2_addition}) and (\ref{eq_2B_Hankel1_addition}). The length of the vector $\bar{\rho}$ is $\rho$, and that of the vector $\bar{\rho}'$ is $\rho'$. The length of the vector $\bar{\rho}-\bar{\rho}'$ (the location of the point of interest measured from the displaced feed) is denoted as $\rho''$. (b) Applying the addition theorem in our structure. The point of interest lies on $\sigma_\text{global}$, and therefore $\rho=R$.}
\label{fig_add}
\end{figure}

A central aim of this paper is to derive the multimodal $S$-matrix of the displaced feed (\ref{eq_2A_Sd_def}) directly from that of the central feed (\ref{eq_2A_Sf_def}) rather than from the complicated MMT.
For this purpose, we employ the addition theorem for Hankel functions of all azimuthal orders:
\begin{equation}
\begin{split}
    &H_m^{(2)}(k_0\rho'')e^{-jm\phi''} = H_m^{(2)}(k_0|\bar{\rho}-\bar{\rho}'|)e^{-jm\phi''} \\
    &=
    \begin{cases}
      \sum_n J_{n-m}(k_0\rho')H_n^{(2)}(k_0\rho)e^{-jn\phi}e^{+j(n-m)\phi'}& \text{if}\ \rho \geq \rho' \\
      \sum_n H_{n-m}^{(2)}(k_0\rho')J_n(k_0\rho)e^{-jn\phi}e^{+j(n-m)\phi'}& \text{if}\ \rho < \rho'
    \end{cases}
\end{split}
\label{eq_2B_Hankel2_addition}
\end{equation}
\begin{equation}
\begin{split}
    &H_m^{(1)}(k_0\rho'')e^{-jm\phi''} = H_m^{(1)}(k_0|\bar{\rho}-\bar{\rho}'|)e^{-jm\phi''} \\
    &=
    \begin{cases}
      \sum_n J_{n-m}(k_0\rho')H_n^{(1)}(k_0\rho)e^{-jn\phi}e^{+j(n-m)\phi'} & \text{if}\ \rho \geq \rho' \\
      \sum_n H_{n-m}^{(1)}(k_0\rho')J_n(k_0\rho)e^{-jn\phi}e^{+j(n-m)\phi'} & \text{if}\ \rho < \rho'
    \end{cases}
\end{split}
\label{eq_2B_Hankel1_addition}
\end{equation}
\noindent where the geometric parameters are illustrated in Fig. \ref{fig_add}(a).
The proof of (\ref{eq_2B_Hankel2_addition}) and (\ref{eq_2B_Hankel1_addition}) can be performed by utilizing raising or lowering operators \cite{Chew_book}, or by matching the large argument approximations of
the Hankel and Bessel functions \cite{Martin_book}.
The first proof is outlined in the Appendix of this paper.
In essence, these addition theorems relate cylindrical waves centered with respect to $(\rho',\phi')$ to those centered
with respect to the global origin.
Therefore, they can help us find the relation between the multimodal $S$-matrices (\ref{eq_2A_Sf_def}) and (\ref{eq_2A_Sd_def}).

To apply these addition theorems to our system, the outgoing propagating electric field, $E_z^+$, along the large cylindrical surface $\sigma_\text{global}$ is studied.
In this case, $\bar{\rho} = \bar{R}$, as shown in Fig. \ref{fig_add}(b).
The field can be expressed in terms of a set of outgoing cylindrical waves centered
with respect to the displaced feed $(\rho', \phi')$,
\begin{equation}
    E_z^+\Big{|}_{\text{along }\sigma_\text{global}}
    = \sum_{m=-M}^{+M} \alpha_{\text{feed},m}^+
    H_m^{(2)}(k_0 |\bar{R}-\bar{\rho'}|) e^{-jm\phi''}.
\label{eq_2B_Ezp_original}
\end{equation}
\noindent Based on our choice of the cylindrical surfaces $\sigma_\text{global}$ and $\sigma_\text{feed}$, $R$ is always greater than $\rho'$.
Consequently, the first part of (\ref{eq_2B_Hankel2_addition}) can be applied to (\ref{eq_2B_Ezp_original})
to yield:
\begin{equation}
\begin{split}
    &E_z^+\Big{|}_{\text{along }\sigma_\text{global}} \\
    &= \sum_m \alpha_{\text{feed},m}^+ \sum_n J_{n-m}(k_0\rho')H_n^{(2)}(k_0 R)e^{-jn\phi}e^{+j(n-m)\phi'} \\
    &= \sum_n \Bigg{[} \sum_m \alpha_{\text{feed},m}^+ J_{n-m}(k_0\rho') e^{+j(n-m)\phi'} \Bigg{]} H_n^{(2)}(k_0 R)e^{-jn\phi}.
\end{split}
\label{eq_2B_Ezp_modified}
\end{equation}
\noindent{Notice that the $H_n^{(2)}(k_0 R)e^{-jn\phi}$ in (\ref{eq_2B_Ezp_modified}) are exactly the outgoing cylindrical waves centered with respect to the global origin and are evaluated at $\sigma_\text{global}$.}
Therefore, the term inside the bracket is actually the corresponding wave amplitudes if we expand $E_z^+$ in terms of these cylindrical waves, as in (\ref{eq_2A_Ez_sigma_global}):
\begin{equation}
    E_z^+\Big{|}_{\text{along }\sigma_\text{global}}
    = \sum_n \alpha_{\text{global},n}^+ H_n^{(2)}(k_0 R)e^{-jn\phi}
\label{eq_2B_Ezp_rewrite}
\end{equation}
\begin{equation}
    \alpha_{\text{global},n}^+ = \sum_m \alpha_{\text{feed},m}^+ J_{n-m}(k_0\rho') e^{+j(n-m)\phi'}.
\label{eq_2B_alpha_relation_sum}
\end{equation}
\noindent{This relationship between the wave amplitudes (\ref{eq_2B_alpha_relation_sum}) can be rewritten in matrix form using the definitions (\ref{eq_2A_alpha_feed_p}) and (\ref{eq_2A_alpha_global_p}),}
\begin{equation}
    \bar{\alpha}^+_\text{global} = \bar{\bar{D}} \bar{\alpha}^+_\text{feed},
\label{eq_2B_alpha_p_relation_matrix}
\end{equation}
\noindent{where information on the feed displacement is embedded in the entries of the $N\times N$ matrix $\bar{\bar{D}}$.
Each entry of $\bar{\bar{D}}$ can be determined by the following equation:}
\begin{equation}
    \bar{\bar{D}} \big[ (M+1)-n, (M+1)-m \big] = J_{n-m}(k_0\rho') e^{+j(n-m)\phi'}.
\label{eq_2B_D_matrix}
\end{equation}
\noindent{The first index in the square bracket indicates the row number of the entry, and the second index indicates the column number.
Note that in $\bar{\alpha}_\text{feed}^+$ (\ref{eq_2A_alpha_feed_p}) and $\bar{\alpha}_\text{global}^+$
(\ref{eq_2A_alpha_global_p}), the entries are arranged in a descending azimuthal order
(the first entry being the $+M$ mode and the last being the $-M$ mode).
The indices in (\ref{eq_2B_D_matrix}) are chosen so that they comply with this descending order.}

A similar approach can be employed to the incoming propagating electric field, $E_z^-$, along $\sigma_\text{global}$.
It can be shown that its wave amplitudes are related in the same manner as (\ref{eq_2B_alpha_p_relation_matrix}):
\begin{equation}
    \bar{\alpha}^-_\text{global} = \bar{\bar{D}} \bar{\alpha}^-_\text{feed}.
\label{eq_2B_alpha_n_relation_matrix}
\end{equation}
\noindent{It is important to examine how (\ref{eq_2B_alpha_p_relation_matrix})-(\ref{eq_2B_alpha_n_relation_matrix}) change when there is no displacement,
i.e. when $\rho'=0$.
If there is no displacement, both expansions (\ref{eq_2A_Ez_sigma_feed}) and (\ref{eq_2A_Ez_sigma_global}) should yield
the same wave amplitudes.
When its argument is zero, the Bessel function of the first kind $J_{n-m}(0)$ only has
a nonvanishing value when $n=m$. In fact, this nonvanishing value is unity.
Hence, when $\rho'=0$, (\ref{eq_2B_D_matrix}) reduces to:}
\begin{equation}
    \bar{\bar{D}} \big[ (M+1)-n, (M+1)-m \big] = \delta_{nm},
\end{equation}
\noindent{which implies that $\bar{\bar{D}}$ becomes an identity matrix.
In other words, $\bar{\alpha}^+_\text{global} = \bar{\alpha}^+_\text{feed}$ and $\bar{\alpha}^-_\text{global} = \bar{\alpha}^-_\text{feed}$, as expected.}

In order to relate the multimodal $S$-matrices (\ref{eq_2A_Sf_def}) and (\ref{eq_2A_Sd_def}), the relationships between power waves rather than those between wave amplitudes are required.
By combining (\ref{eq_2A_A1feed_def}), (\ref{eq_2A_A1_def}), and (\ref{eq_2B_alpha_p_relation_matrix}), $\bar{A}^{(1)}$ can be neatly expressed in terms of $\bar{A}^{(1)}_\text{feed}$:
\begin{equation}
\begin{split}
    \bar{A}^{(1)} &=
    \Big{[} \bar{\bar{c}}^+_R \bar{\bar{H}}^+_R \bar{\bar{D}} (\bar{\bar{H}}^+_r)^{-1} (\bar{\bar{c}}^+_r)^{-1} \Big{]} \bar{A}^{(1)}_\text{feed} \\
    &\triangleq \bar{\bar{D}}_A \bar{A}^{(1)}_\text{feed}.
\end{split}
\label{eq_2B_PowerWaveA}
\end{equation}
\noindent{Similarly, we have for $\bar{B}^{(1)}$ and $\bar{B}^{(1)}_\text{feed}$,}
\begin{equation}
\begin{split}
    \bar{B}^{(1)} &=
    \Big{[} \bar{\bar{c}}^-_R \bar{\bar{H}}^-_R \bar{\bar{D}} (\bar{\bar{H}}^-_r)^{-1} (\bar{\bar{c}}^-_r)^{-1} \Big{]} \bar{B}^{(1)}_\text{feed} \\
    &\triangleq \bar{\bar{D}}_B \bar{B}^{(1)}_\text{feed}.
\end{split}
\label{eq_2B_PowerWaveB}
\end{equation}
\noindent Based on the definitions (\ref{eq_2B_PowerWaveA}) and (\ref{eq_2B_PowerWaveB}), the dimensions of the matrices $\bar{\bar{D}}_A$ and $\bar{\bar{D}}_B$ are both $N\times N$.

Now that the relationships between power waves have been found, the multimodal $S$-matrix of a displaced feed can also be derived. By applying (\ref{eq_2B_PowerWaveA}), (\ref{eq_2B_PowerWaveB}) and rewriting (\ref{eq_2A_Sd_def}), we obtain:
\begin{equation}
\begin{split}
    &\begin{bmatrix}
    \bar{\bar{I}}_{1\times 1} & \bar{\bar{O}}_{1\times N} \\ \bar{\bar{O}}_{N\times 1} & \bar{\bar{D}}_A
    \end{bmatrix}
    \begin{bmatrix}
    \bar{B}^{(0)} \\ \bar{A}^{(1)}_\text{feed}
    \end{bmatrix}
    \\
    &=
    \begin{bmatrix}
    \bar{\bar{S}}_{d,11} & \bar{\bar{S}}_{d,12} \\ \bar{\bar{S}}_{d,21} & \bar{\bar{S}}_{d,22}
    \end{bmatrix}
    \cdot
    \begin{bmatrix}
    \bar{\bar{I}}_{1\times 1} & \bar{\bar{O}}_{1\times N} \\ \bar{\bar{O}}_{N\times 1} & \bar{\bar{D}}_B
    \end{bmatrix}
    \begin{bmatrix}
    \bar{A}^{(0)} \\ \bar{B}^{(1)}_\text{feed}
    \end{bmatrix},
\end{split}
\label{eq_2B_derivation1}
\end{equation}
\noindent where $\bar{\bar{I}}_{1\times 1}$, $\bar{\bar{O}}_{1\times N}$, and $\bar{\bar{O}}_{N\times 1}$ denote the $1\times 1$ identity matrix, $1\times N$ zero matrix, and $N\times 1$ zero matrix, respectively.
After rearranging (\ref{eq_2B_derivation1}) and comparing the result
with the definition (\ref{eq_2A_Sf_def}), the following equation is derived:
\begin{equation}
\begin{split}
    &\begin{bmatrix}
    \bar{\bar{S}}_{d,11} & \bar{\bar{S}}_{d,12} \\ \bar{\bar{S}}_{d,21} & \bar{\bar{S}}_{d,22}
    \end{bmatrix}
    \\
    &=
    \begin{bmatrix}
    \bar{\bar{I}}_{1\times 1} & \bar{\bar{O}}_{1\times N} \\ \bar{\bar{O}}_{N\times 1} & \bar{\bar{D}}_A
    \end{bmatrix}
    \begin{bmatrix}
    \bar{\bar{S}}_{f,11} & \bar{\bar{S}}_{f,12} \\ \bar{\bar{S}}_{f,21} & \bar{\bar{S}}_{f,22}
    \end{bmatrix}
    \begin{bmatrix}
    \bar{\bar{I}}_{1\times 1} & \bar{\bar{O}}_{1\times N} \\ \bar{\bar{O}}_{N\times 1} & \bar{\bar{D}}_B
    \end{bmatrix}^{-1},
\end{split}
\label{eq_2B_derivation2}
\end{equation}

\noindent{or equivalently:}
\begin{equation}
\begin{split}
    \bar{\bar{S}}_{d,11} &= \bar{\bar{S}}_{f,11} \\
    \bar{\bar{S}}_{d,12} &= \bar{\bar{S}}_{f,12} \bar{\bar{D}}_B^{-1} \\
    \bar{\bar{S}}_{d,21} &= \bar{\bar{D}}_A \bar{\bar{S}}_{f,21} \\
    \bar{\bar{S}}_{d,22} &= \bar{\bar{D}}_A \bar{\bar{S}}_{f,22} \bar{\bar{D}}_B^{-1},
\end{split}
\label{eq_2B_result}
\end{equation}
\noindent{where the inversion of block diagonal matrices \cite{Lu_BlockInversion} is utilized. It is clear from (\ref{eq_2B_result}) that the multimodal $S$-matrix of a displaced feed (\ref{eq_2A_Sd_def}) can be easily obtained from that of the central feed (\ref{eq_2A_Sf_def}).
The cumbersome MMT only needs to be conducted once to calculate (\ref{eq_2A_Sf_def}), and any displacement of the coaxial feed can be taken into account through simple matrix multiplication.
Consequently, a major goal of this paper has been achieved.}


With the scattering information of the displaced feed, a synthesis or design problem can
now be tackled. The interaction between this displaced feed and the cascaded cylindrical metasurfaces
can be taken into account by solving (\ref{eq_2A_Sd_def}) and (\ref{eq_2A_Sm_def}) jointly.
Consider the setup in Fig. \ref{fig_S_def} again.
In theory, there is no reflection from infinity, so we set $\bar{B}^{(2)}=\bar{0}$ which is a zero vector.
The power waves in the central region can be expressed in terms of the incident power
waves from the coaxial feed, $\bar{A}^{(0)}$, as:
\begin{equation}
    \bar{A}^{(1)} = (\bar{\bar{I}}_{N\times N}- \bar{\bar{S}}_{d,22}\bar{\bar{S}}_{M,11})^{-1} \bar{\bar{S}}_{d,21} \bar{A}^{(0)}
\label{eq_2B_interaction_A1}
\end{equation}
\begin{equation}
    \bar{B}^{(1)} = \bar{\bar{S}}_{M,11}(
                    \bar{\bar{I}}_{N\times N}- \bar{\bar{S}}_{d,22}\bar{\bar{S}}_{M,11})^{-1} \bar{\bar{S}}_{d,21} \bar{A}^{(0)},
\label{eq_2B_interaction_B1}
\end{equation}
\noindent{where $\bar{\bar{I}}_{N\times N}$ denotes an $N\times N$ identity matrix. By substituting (\ref{eq_2B_interaction_A1}) and (\ref{eq_2B_interaction_B1}) into (\ref{eq_2A_Sd_def}), the reflected power wave in the coaxial feed is calculated as,}
\begin{equation}
\begin{split}
    &\bar{B}^{(0)}
    \\
    &= \Big{[} \bar{\bar{S}}_{d,11} +
    \bar{\bar{S}}_{d,12}\bar{\bar{S}}_{M,11}(
    \bar{\bar{I}}_{N\times N}- \bar{\bar{S}}_{d,22}\bar{\bar{S}}_{M,11})^{-1} \bar{\bar{S}}_{d,21}
    \Big{]} \bar{A}^{(0)},
\end{split}
\label{eq_2B_interaction_B0}
\end{equation}
\noindent{from which the reflection coefficient of the device can be deduced. Alternatively, (\ref{eq_2A_Sm_def}) yields the power wave in the outer region,}
\begin{equation}
    \bar{A}^{(2)} = \bar{\bar{S}}_{M,21}(\bar{\bar{I}}_{N\times N}- \bar{\bar{S}}_{d,22}\bar{\bar{S}}_{M,11})^{-1} \bar{\bar{S}}_{d,21} \bar{A}^{(0)}.
\label{eq_2B_interaction_A2}
\end{equation}
\noindent{Finally, arbitrary field transformations from this displaced coaxial feed are accomplished
by matching this power wave $\bar{A}^{(2)}$ to the desired or stipulated field through an optimization algorithm.}  


\section{Validation Examples}
\label{sec:examples}
An innovative antenna beam-shaping shell is demonstrated in this section.
The results not only verify the presented theory, but also demonstrate its effectiveness.
Moreover, an impedance matching structure is introduced and incorporated into the device
to enhance its practical performance characteristics.

\subsection{The Antenna Beam-Shaping Shell}

A beam-shaping shell is developed that converts the input field originating from a 
displaced coaxial feed to a narrow radiated beam.
Although the functionality of the resulting device is similar to the azimuthally-symmetric case presented in \cite{my_EuCAP_2023}, the aim herein is to accomplish the design with azimuthally-varying cylindrical metasurfaces.
The additional degrees of freedom introduced by azimuthal variation allows for a design with 
fewer cylindrical metasurface layers. 
Furthermore, they also help achieve a narrower beam and/or a lower reflection coefficient.

Our goal in this design example is to synthesize a narrow beam radiated in the $\phi=0^\circ$ direction by tailoring cylindrical electromagnetic waves.
This idea was inspired by \cite{Ziolkowski_PRL_2018, Ziolkowski_OJAP_2022} where dielectric shells 
with large positive and negative relative permittivities were applied to control the cylindrical waves emitted by a line source.
Dielectric materials with large negative relative permittivities are difficult to realize.
Therefore, cylindrical metasurfaces consisting of patterned metallic claddings on commercial dielectric substrates are utilized to control the radiated waves.
Their use furnishes the important advantage of circumventing the need for dielectric materials with extreme relative permittivities.

As presented in \cite{Ziolkowski_PRL_2018, Ziolkowski_OJAP_2022}, the Dirac delta function $\delta(\phi)$ describes a radiation pattern with an idealized (infinitely narrow) beam pointing solely in the forward ($\phi=0^\circ$) direction. It can be expressed in terms of azimuthal modes based on Fourier series, where the coefficient associated with each mode is identical:
\begin{equation}
    \delta(\phi) = \frac{1}{2\pi}\sum_{m=-\infty}^{+\infty}e^{-jm\phi}.
\label{eq_3A_delta_fx}
\end{equation}
\noindent{Therefore, we set our target electric field in the outer region of the 
cylindrical metasurfaces to be:}
\begin{equation}
    E_{z,\text{target}}(\rho,\phi) = A \sum_{m=-\infty}^{+\infty} (-j)^mH_m^{(2)}(k_0\rho)e^{-jm\phi},
\label{eq_3A_target}
\end{equation}
\noindent{where $(-j)^m$ is the corresponding wave amplitude of the cylindrical wave with azimuthal order $m$, and $A$ is an arbitrary constant.
Note that only Hankel functions of the second kind are present in (\ref{eq_3A_target})
since it is assumed that there are no incoming cylindrical waves from infinity.
The reason for setting the target field to (\ref{eq_3A_target}) can be understood immediately from its 
far-field expression. In particular, the large argument approximation of the Hankel functions can be employed
in the far-field where $k_0\rho \gg 1$, i.e.,
\cite{Harrington_book}:}
\begin{equation}
    H_m^{(2)}(k_0\rho) \approx \sqrt{\frac{2j}{\pi k_0\rho}}(+j)^m e^{-jk_0\rho}.
\label{eq_3A_Hankel_approx}
\end{equation}
\noindent{Consequently, (\ref{eq_3A_target}) becomes,}
\begin{equation}
    E_{z,\text{target}}(\rho,\phi) \approx A \sqrt{\frac{2j}{\pi k_0\rho}} e^{-jk_0\rho} \sum_{m=-\infty}^{+\infty} e^{-jm\phi}.
\label{eq_3A_target_ff}
\end{equation}
\noindent{It is clear that the expansion coefficient associated with each azimuthal mode in (\ref{eq_3A_target_ff}) is identical, as it is in the Dirac delta function (\ref{eq_3A_delta_fx}). Thus,
the desired field in the outer region described by (\ref{eq_3A_target}) yields an idealized (infinitely narrow) radiated beam.
In practice, it is impossible to control an infinite number of azimuthal modes.
Hence, (\ref{eq_3A_delta_fx}) and (\ref{eq_3A_target}) need to be truncated.
This results in a narrow radiated beam of finite beamwidth in the forward direction.}

The operating frequency of our example beam-shaping shell is set at $f=10$ GHz. 
The displaced feed is located at $(\rho',\phi')=(0.8\lambda,0)$ where $\lambda$ denotes the operating wavelength 
in free-space.
The detailed dimensions of the coaxial cable are $a=0.45$ mm, $b=1.5$ mm (see Fig. \ref{fig_S_def}), and the dielectric within the cable is Teflon with $\varepsilon=2.2 \, \varepsilon_0$, making the characteristic impedance of the coaxial cable $Z_\text{coax}=50$ $\Omega$.
A single layer of cylindrical metasurface, whose radius is $\rho_\text{MTS}=2.7\lambda$, is centered with respect to the global origin and placed within a radial waveguide whose height is $h=5$ mm.
The physical dimensions of the coaxial cable, as well as the radial waveguide, satisfy (\ref{eq_2A_h_limitation}) and (\ref{eq_2A_ab_limitation}). Therefore, the analysis developed in the previous section is 
a valid approach to evaluate the performance of the example structure.
Furthermore, they are also identical to those adopted in \cite{my_AWPL_2023}.
Hence, the multimodal $S$-matrix of the coaxial-waveguide junction (\ref{eq_2A_Sf_def}) is the same as the one  
obtained in that work.

In multimodal wave matrix theory, a cylindrical metasurface can be described mathematically by a $\phi$-dependent admittance profile $Y_\text{MTS}(\phi)$ which relates the induced surface current density to the averaged electric field \cite{my_TAP_2023}.
For a lossless design, $Y_\text{MTS}(\phi)$ is purely imaginary and can be expressed using the susceptance $Y_\text{MTS}(\phi) = jB_\text{MTS}(\phi)$. 
The susceptance profile of the cylindrical metasurface in our solver was optimized to have 
the electric field in the outer region, which is related to the power wave $\bar{A}^{(2)}$ given by (\ref{eq_2B_interaction_A2}), come as close to the desired field as possible. The target field was selected to consist of the first 11 terms (from $m=-5$ to $m=+5$) of the infinite series (\ref{eq_3A_target}). 
In order to avoid highly-oscillatory susceptance profiles which are difficult to realize, we have enforced a smooth, sinusoidal azimuthal variation of $B_\text{MTS}(\phi)$ in the multimodal wave matrix theory.

\begin{figure}[!t]
\centering
\subfloat[\footnotesize]{%
    \includegraphics[width=0.42\linewidth]{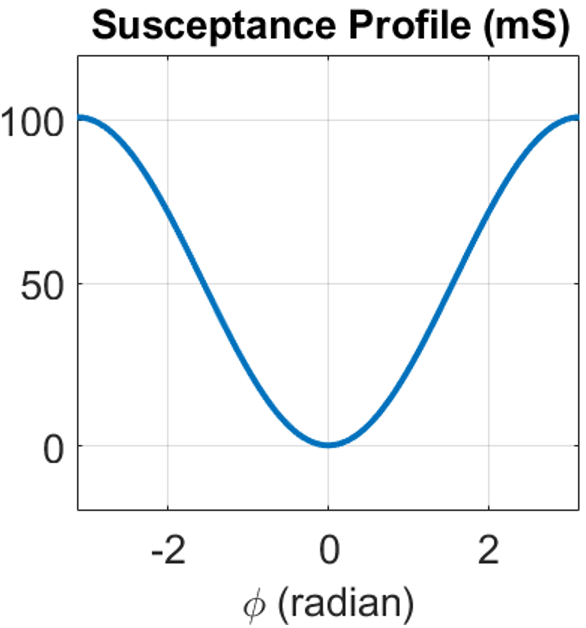}}
    \hfill
\subfloat[\footnotesize]{%
    \includegraphics[width=0.56\linewidth]{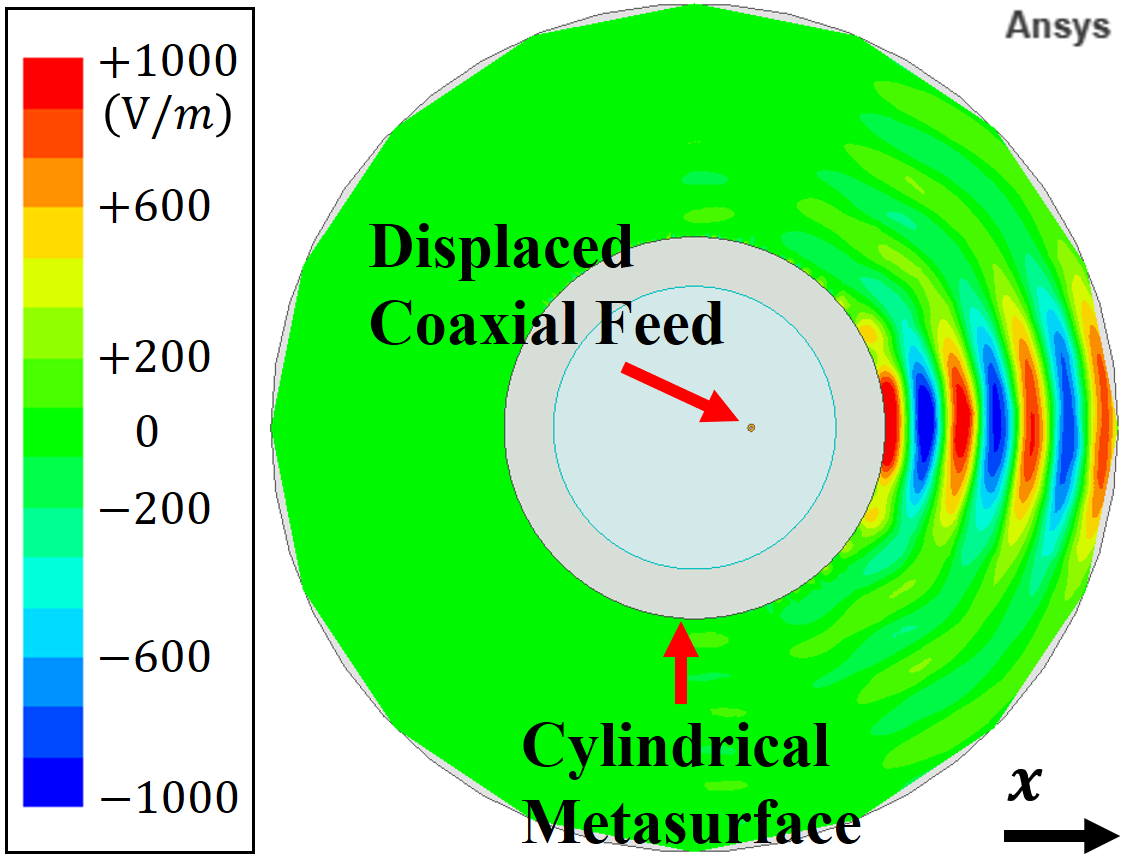}}
\caption{The optimized beam-shaping shell. (a) The susceptance profile $B_\text{MTS}(\phi)$ of the cylindrical metasurface located at $\rho_\text{MTS}=2.7\lambda$. The admittance profile is given by $Y_\text{MTS}(\phi) = jB_\text{MTS}(\phi)$. (b) The real part of the simulated electric field $E_z$ in the outer region (a time-snapshot of the $E_z$ field for $0^\circ$ phase in the period $T = 1/f$).}
\label{fig_example_initial}
\end{figure}

The synthesized results of the beam-shaping shell are shown in Fig. \ref{fig_example_initial}.
The optimized susceptance profile $B_\text{MTS}(\phi)$ of the single-layer cylindrical metasurface, illustrated in Fig. \ref{fig_example_initial}(a), is discretized into 60 unit cells and implemented as impedance boundary conditions in Ansys HFSS.
The smooth azimuthal variation of the susceptance profile $B_\text{MTS}(\phi)$ indicates another advantage of our formulation: extreme discretization is not required.
The simulated performance of the device shown in Fig. \ref{fig_example_initial}(b) clearly demonstrates a narrow radiation beam in the $\phi=0^\circ$ direction as specified.
Our analysis method predicted a reflection coefficient of $0.73\angle 11.75^\circ$ 
at the $z=0$ plane of the coaxial feed in Fig. \ref{fig_S_def},   
while the full-wave simulation yielded a value of $0.75\angle 13.50^\circ$.
These very close values verify its accuracy.

A related characteristic of interest is the 2-D directivity (recall that the field in the radial waveguide is a 2-D one). It is calculated based on the following definition \cite{Ziolkowski_PRL_2018, Ziolkowski_OJAP_2022}:
\begin{equation}
    D_{2D}(\phi=0^\circ)
    = \frac{2\pi\rho \big{|} E_{z,ff}(\phi=0^\circ) \big{|}^2}{\int_0^{2\pi} \big{|} E_{z,ff}(\phi) \big{|}^2 \rho d\phi}.
\label{eq_3A_D2D_def}
\end{equation}
\noindent{If the electric field in the far-field is expressed in terms of azimuthal modes with corresponding coefficients $C_m$:}
\begin{equation}
    E_{z,ff}(\phi) = \sum_{m=-\infty}^{+\infty} C_m e^{-jm\phi},
\label{eq_3A_D2D_expression}
\end{equation}
\noindent{then the numerator of (\ref{eq_3A_D2D_def}) can be calculated knowing that,}
\begin{equation}
    E_{z,ff}(\phi=0^\circ) = \sum_{m=-\infty}^{+\infty} C_m,
\label{eq_3A_D2D_numer1}
\end{equation}
\noindent and, hence, that
\begin{equation}
    \Big{|} E_{z,ff}(\phi=0^\circ) \Big{|}^2 = \Bigg{|} \sum_{m=-\infty}^{+\infty} C_m \Bigg{|}^2.
\label{eq_3A_D2D_numer2}
\end{equation}
\noindent{Moreover, the integrand of the denominator in (\ref{eq_3A_D2D_def}) can be written as,}
\begin{equation}
    \Big{|} E_{z,ff}(\phi) \Big{|}^2 = \sum_{m=-\infty}^{+\infty} \sum_{m'=-\infty}^{+\infty} C_m(C_{m'})^* e^{-j(m-m')\phi}.
\label{eq_3A_D2D_denom1}
\end{equation}
\noindent{After integration, only the terms with $m=m'$ remain yielding:}
\begin{equation}
    \int_0^{2\pi} \Big{|} E_{z,ff}(\phi) \Big{|}^2 d\phi = \sum_{m=-\infty}^{+\infty} |C_m|^2.
\label{eq_3A_D2D_denom2}
\end{equation}
\noindent{Therefore, the 2-D directivity (\ref{eq_3A_D2D_def}) is readily calculated as:}
\begin{equation}
    D_{2D}(\phi=0^\circ) = \frac{|\sum_m C_m|^2}{\sum_m |C_m|^2}.
\label{eq_3A_D2D_result}
\end{equation}

\begin{figure}[!t]
\centerline{\includegraphics[height=7.6cm]{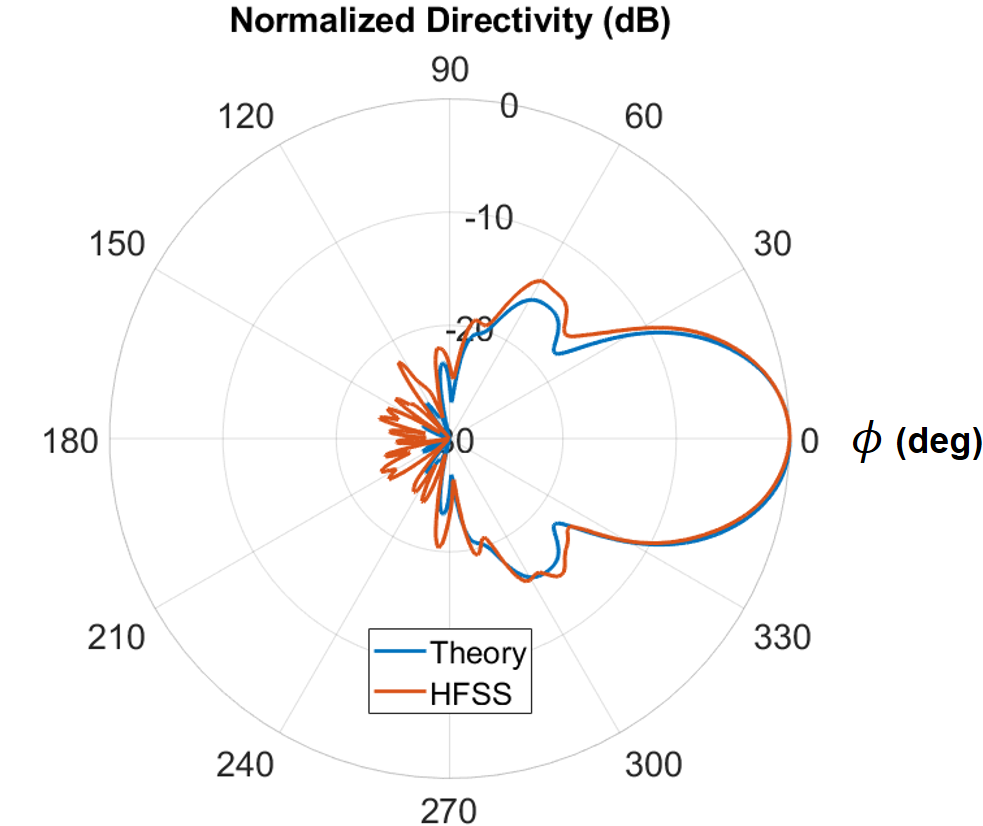}}
\caption{Radiation pattern (normalized 2-D directivity) of the beam-shaping system in its $\theta=90^\circ$ plane. Both the theoretically predicted and the full-wave simulated patterns are shown.}
\label{fig_example_RadPattern}
\end{figure}

Since only 11 terms in the series (\ref{eq_3A_target}) representing the radiated field were considered, 
the highest 2-D directivity that could have been obtained is 11.0 or $10.41$ dB.
With the susceptance profile shown in Fig. \ref{fig_example_initial}(a), our beam-shaping system produces a 2-D directivity of 10.16 dB, while the corresponding full-wave simulation results in \ref{fig_example_initial}(b) 
attained a value of $10.14$ dB.
Additionally, the radiation patterns in the $\theta=90^\circ$ plane calculated from both the theory and the HFSS simulation are plotted and compared in Fig. \ref{fig_example_RadPattern}.
Close agreement between the two curves is evident.
The small lobes in the left portion of the full-wave simulated pattern have been confirmed to be due to the level of discretization of the mesh employed throughout the metasurface and overall structure.  
The 3-dB beamwidth of the main lobe produced by the designed beam-shaping shell is approximately $30^\circ$ and the side lobe level is $-13.8$ dB. 
Moreover, the device also offers a very high front-to-back ratio that is greater than $25$ dB. 
This example clearly illustrates the capability of the developed design framework to achieve arbitrary field transformations from a metasurface-based structure excited by a displaced coaxial feed. 

\subsection{The Impedance Matching Structure}

Another important figure of merit of practical importance is the input impedance of our structure with a realistic coaxial feed and impedance matching element.
Previous researchers have proposed several methods to improve matching to a coaxial feed.
A top-loading disk (metallic puck) was inserted into the junction between the coaxial cable and the radial waveguide
in \cite{Shen_MOTL_1999}.
However, this method significantly modifies the feeding structure and, hence, necessitates a recalculation of the MMT-based scattering properties \cite{my_AWPL_2023, my_APS_2022, Shen_MOTL_1999, Heebl_PRA_2016, Faris_OJAP_2021, Eleftheriades_TMTT_1994, Kuhn_AEU_1973}.
Another matching technique is to design a quarter-wave impedance transformer within the coaxial cable \cite{Pozar_book}.
Since the impedance transformer is physically separated from the coaxial-waveguide junction, it does not disturb the feeding structure and, therefore, can be applied easily to the device \cite{Ettorre_TAP_2012}.

\begin{figure}[!t]
\centering
\subfloat[\footnotesize]{%
    \includegraphics[width=0.5\linewidth]{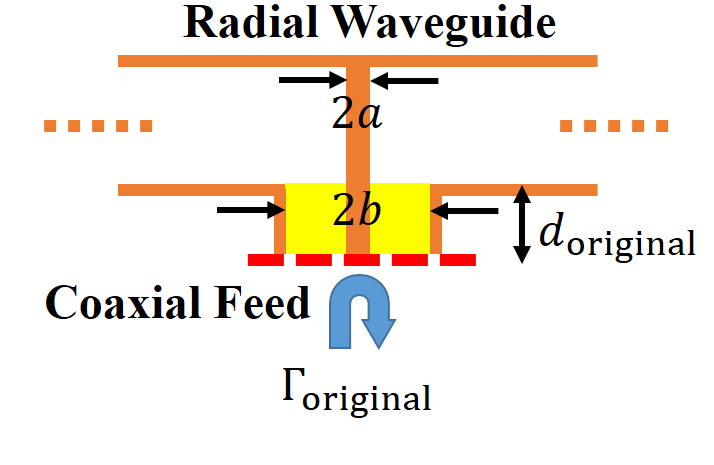}}
\subfloat[\footnotesize]{%
    \includegraphics[width=0.5\linewidth]{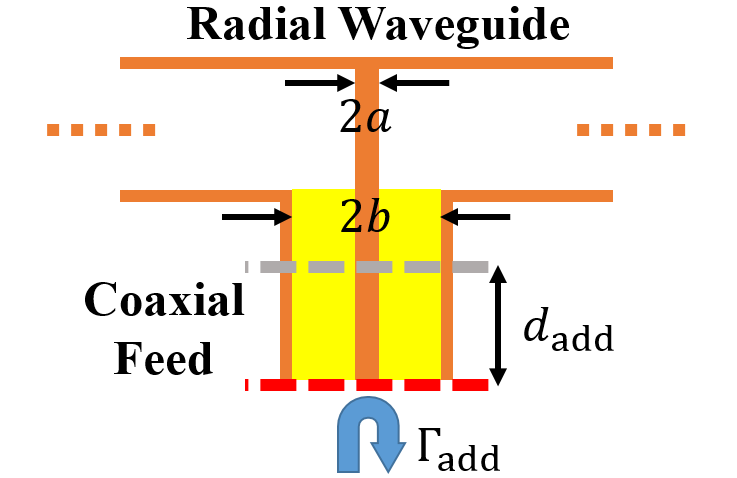}}
\\
\subfloat[\footnotesize]{%
    \includegraphics[width=\linewidth]{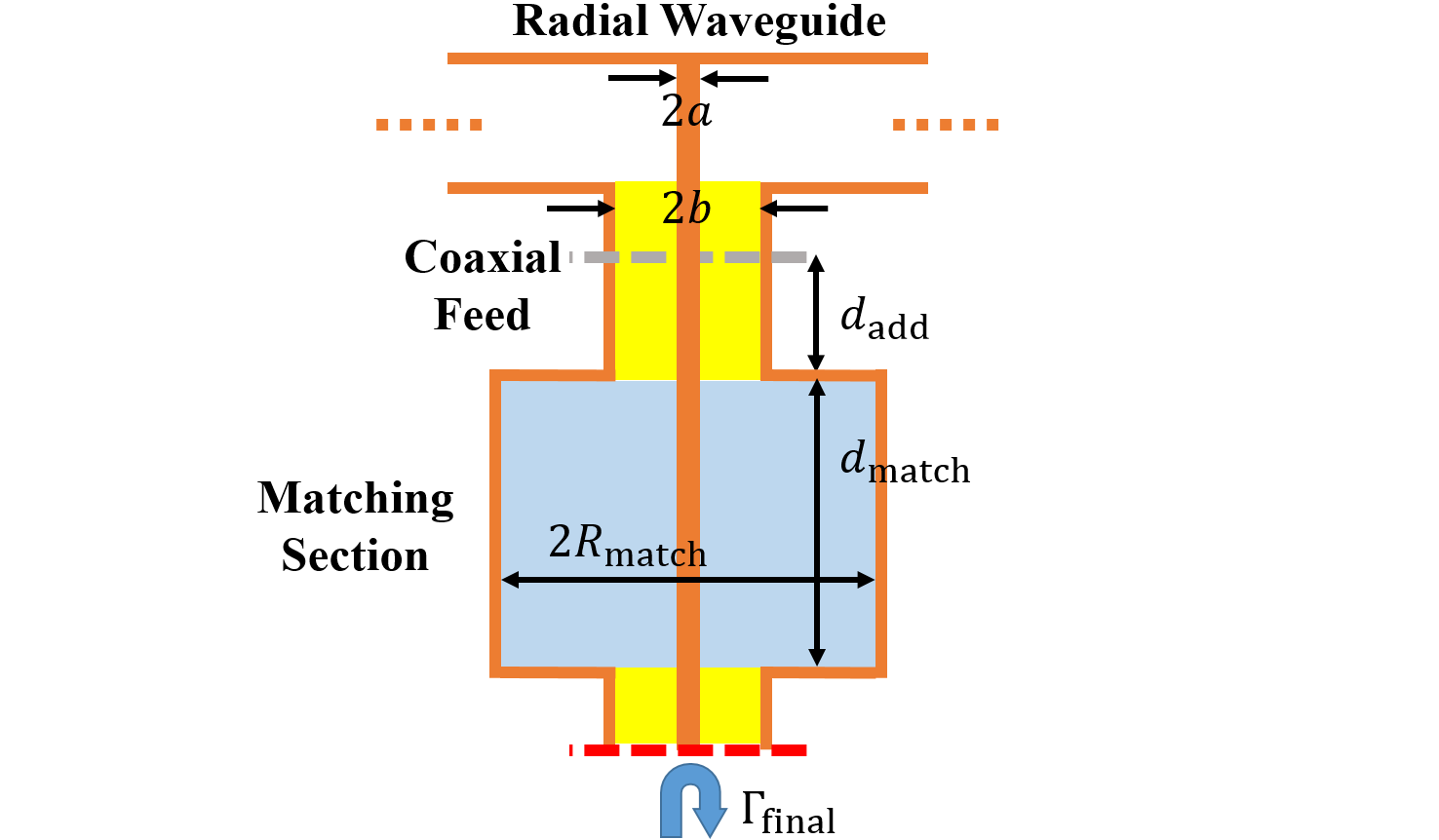}}
\caption{Designing the matching section for the displaced coaxial feed. (a) The cross-sectional view of the original coaxial-waveguide junction. (b) Extending the feed line to make the input impedance real.
(c) Adding a quarter-wave impedance transformer to attain a final input impedance of $50$ $\Omega$.}
\label{fig_example_matching}
\end{figure}

The design evolution of the coaxial quarter-wave impedance transformer for our beam-shaping device is illustrated in the cross-sectional views given in Fig. \ref{fig_example_matching}.
The original structure in Fig. \ref{fig_example_matching}(a) is a simple junction between the coaxial cable and the radial waveguide. It has a reflection coefficient denoted by $\Gamma_\text{original}$.
With our choice of the reference surface for the coaxial feed, explained in (\ref{eq_2A_Sf_def}),
no extra length of coax is present, and we characterize this fact with $d_\text{original}=0$.
Next, the coaxial cable is lengthened by $d_\text{add}$, as illustrated in Fig. \ref{fig_example_matching}(b), in order to transform the reflection coefficient at its end, $\Gamma_\text{add}$, and the corresponding input impedance, $Z_\text{add}$, to real quantities.
They are calculated as:
\begin{equation}
    \Gamma_\text{add} = \Gamma_\text{original}\cdot e^{-2j(\omega\sqrt{\mu_0\varepsilon})d_\text{add}},
\label{eq_3B_Gamma_add}
\end{equation}
\begin{equation}
    Z_\text{add} = Z_\text{coax}\frac{1+\Gamma_\text{add}}{1-\Gamma_\text{add}} = 50\frac{1+\Gamma_\text{add}}{1-\Gamma_\text{add}} \ (\Omega).
\label{eq_3B_Z_add}
\end{equation}
\noindent{Finally, quarter-wave matching can be performed with the real $Z_\text{add}$ to minimize the reflection coefficient of the device, $\Gamma_\text{final}$.
It was achieved by introducing a modified coaxial cable section as shown in Fig. \ref{fig_example_matching}(c), then carefully selecting the dielectric permittivity within it, $\varepsilon_\text{match}$, and optimizing the radius
of its outer conductor, $R_\text{match}$, according to the resulting characteristic impedance $Z_\text{match}$:}
\begin{equation}
    Z_\text{match}^2 = Z_\text{coax}Z_\text{add}.
\label{eq_3B_Z_match}
\end{equation}
\noindent{The length of this matching section, $d_\text{match}$, has to be a quarter of the operating wavelength in the presence of the dielectric filling, i.e., $\lambda/(4\sqrt{\varepsilon_\text{match}/\varepsilon_0})$, to achieve a perfect match.}

This procedure was adopted to design an impedance matching structure for the example beam-shaping shell device.
The original reflection coefficient was $\Gamma_\text{original}=0.75\angle 13.50^\circ$. The smallest length that renders a real $\Gamma_\text{add}$ (\ref{eq_3B_Gamma_add}) is $d_\text{add} = 0.38$ mm.
Nevertheless, this value is not feasible in practice because the resulting matching section will be almost directly connected to the coax-waveguide junction.
The resulting evanescent waves generated from the discontinuities of this short matching section
would significantly impact the characterization of the coaxial feed.
Hence, $d_\text{add}$ was selected to be the next possible value, $10.5$ mm, that yield a real $\Gamma_\text{add}$.
This extra section of coaxial cable has a real reflection coefficient $\Gamma_\text{add}=0.75$ (or $-2.5$ dB)
and an input impedance $Z_\text{add}=350\ \Omega$. Consequently, the required characteristic impedance of the
matching section, given by (\ref{eq_3B_Z_match}), is $Z_\text{match}\approx132.29\ \Omega$.
The dielectric material within the matching section of the quarter-wave impedance transformer was selected to be free-space (air). The theoretical formulas (\ref{eq_3B_Gamma_add})-(\ref{eq_3B_Z_match}) then gave the required dimensions $d_\text{match}=7.5$ mm and $R_\text{match}=4.1$ mm, as listed in Table \ref{tab1}.
With these dimensions, the magnitude of the final reflection coefficient, $\Gamma_\text{final}=0.198$ (or $-14.1$ dB) still deviated from a perfect match. This was due to the parasitic effects stemming from the large difference between $R_\text{match}$ and the coaxial outer radius $b$.

In order to mitigate these parasitic effects, the dimensions $R_\text{match}$ and $d_\text{match}$ were optimized. As shown in Table \ref{tab1}, the subsequent optimized dimensions $R_\text{match} = 3.3$ mm and $d_\text{match}=7.0$ mm resulted in a reflection coefficient of $\Gamma_\text{final}=0.088$ (or $-21.1$ dB), significantly lower than without the matching network.

\begin{figure}[!t]
\centerline{\includegraphics[width=\linewidth]{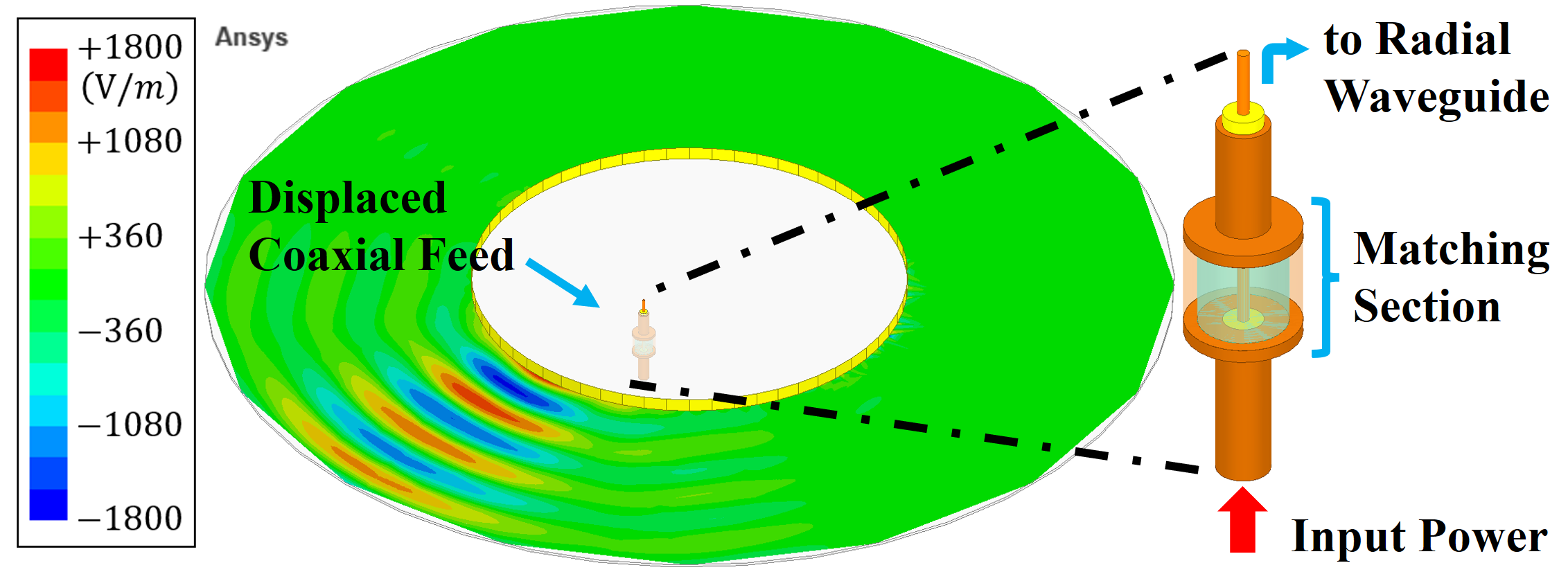}}
\caption{The beam-shaping shell system together with a perspective view of its coaxial feed,
which now includes the quarter-wave impedance transformer matching section illustrated in Fig. \ref{fig_example_matching}.
The dimensions of the impedance matching structure are detailed in Table \ref{tab1}. The real part of the simulated electric field $E_z$ in the outer region is shown.}
\label{fig_example_modified}
\end{figure}

\begin{figure}[!t]
\centerline{\includegraphics[height=7.6cm]{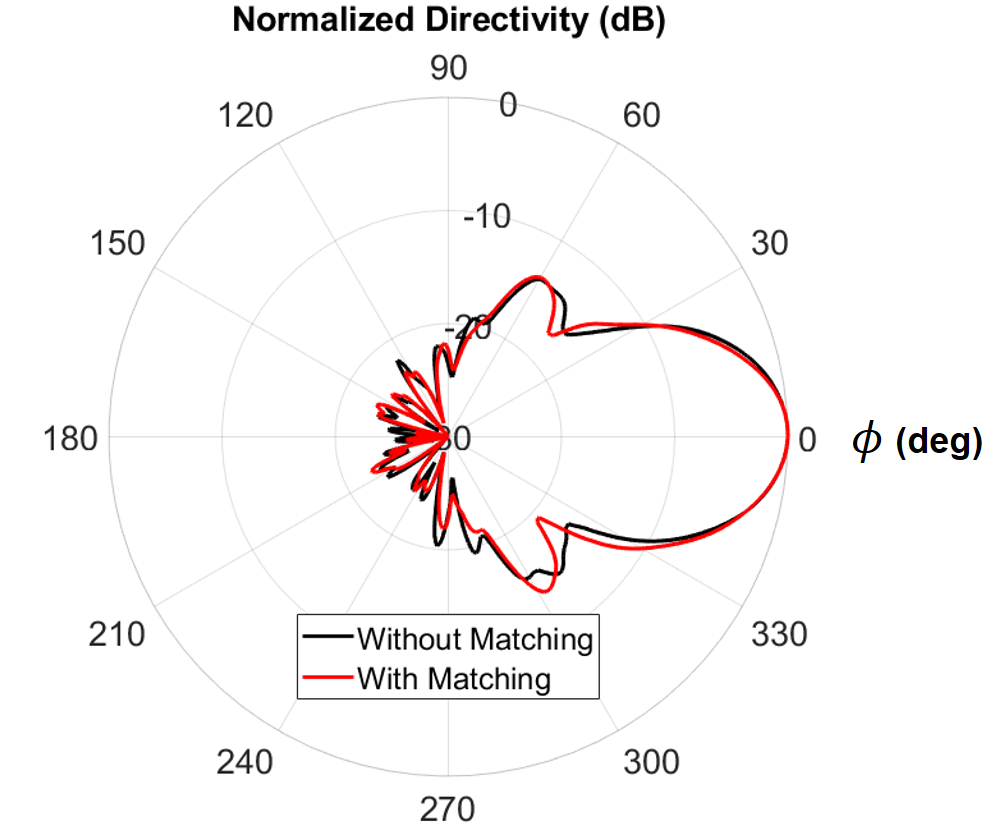}}
\caption{Radiation (normalized 2-D directivity) patterns in the $\theta=90^\circ$ plane of the beam-shaping shell device with and without the optimized impedance matching structure.}
\label{fig_example_RadPattern_modified}
\end{figure}

This optimized impedance matching structure was applied to the example beam-shaping shell system.
Fig. \ref{fig_example_modified} illustrates the full-wave simulations obtained with Ansys HFSS.
Fig. \ref{fig_example_RadPattern_modified} compares the radiation pattern of the original device to that of the impedance matched version.
The quarter-wave transformer greatly improved impedance match of the device with only a negligible disturbance in its radiation pattern.

\begin{table}[!t]
\caption{Design of the Quarter-Wave Impedance Transformer}
\center
\setlength{\tabcolsep}{8pt}
\begin{tabular}{|c|c|c|}
\hline
& Initial Design Based on Formulas & Optimized Design \\
\hline
$d_\text{match}$ & 7.5 (mm) & 7.0 (mm) \\
\hline
$R_\text{match}$ & 4.1 (mm) & 3.3 (mm) \\
\hline
$\varepsilon_\text{match}$ & $\varepsilon_0$ (air) & $\varepsilon_0$ (air) \\
\hline
$\Gamma_\text{final}$ & 0.198 & 0.088 \\
\hline
\end{tabular}
\label{tab1}
\end{table}

\section{Conclusion}
\label{sec:conclusion}

In this paper, a rigorous investigation of a set of layered, concentric cylindrical metasurfaces excited by a displaced, off-center coaxial feed in a radial waveguide was presented.
By employing the Hankel function addition theorem for all azimuthal orders,
the scattering properties of the displaced coaxial feed were obtained efficiently from those of a central coaxial feed.
The formulation was used to derive the multimodal $S$-matrix of the displaced coaxial feed.
It was then combined with multimodal wave matrix theory, which has been used successfully to model
concentric metasurfaces, to realize the design of an arbitrary field transforming structure excited by a
displaced coaxial feed.
The modeling approach was verified with a beam-shaping shell device which generated a specified narrow radiated beam.
Excellent agreement between theoretical predictions and the corresponding full-wave simulations showcased the accuracy of the developed analysis and synthesis method.

Moreover, the practical performance of this example device was significantly enhanced with the design and seamless integration of a quarter-wave impedance transformer into the displaced coaxial feed.
This outcome demonstrates that characterizing a realistic coaxial feed not only provides accurate modeling of the feed-metasurface interaction, but it also allows for the design of fully-integrated matching feed networks.
Future work includes the characterization of multiple coaxial feeds to realize advanced cylindrical-metasurface-based MIMO devices.

\appendices

\section{Addition Theorem for Hankel Functions of Any Azimuthal Order}

The addition theorem for Hankel functions of all azimuthal orders, given by (\ref{eq_2B_Hankel2_addition}) and (\ref{eq_2B_Hankel1_addition}), is derived in detail in this appendix.
The proof is based upon the raising and lowering operators discussed in \cite{Chew_book} applied to
elementary cylindrical waves.
Specifically, the raising operator $\hat{R}_{(x,y)}$ is defined as:
\begin{equation}
    \hat{R}_{(x,y)} =-\frac{1}{k}\Big{(} \frac{\partial}{\partial x} - j\frac{\partial}{\partial y} \Big{)},
\label{eq_App_Raising_def}
\end{equation}
\noindent{where $x=\rho\cos\phi$ and $y=\rho\sin\phi$ represent the rectangular coordinates
corresponding to the cylindrical coordinate $(\rho, \phi)$. By applying (\ref{eq_App_Raising_def}) on an elementary cylindrical wave of order $n$, which depends on both the radial and azimuthal coordinates,} the azimuthal order of the wave is raised by exactly 1, i.e.,
\begin{equation}
    \hat{R}_{(x,y)} \big{[}B_n(k_0\rho)e^{-jn\phi}\big{]} = B_{n+1}(k_0\rho)e^{-j(n+1)\phi},
\label{eq_App_Raising_operation}
\end{equation}
\noindent{where $B_n$ denotes the solution of the Bessel differential equation of order $n$.
Similarly, a lowering operator $\hat{L}_{(x,y)}$ can be defined such that it lowers the order of an
elementary cylindrical waves by 1:}
\begin{equation}
    \hat{L}_{(x,y)} =+\frac{1}{k}\Big{(} \frac{\partial}{\partial x} + j\frac{\partial}{\partial y} \Big{)},
\label{eq_App_Lowering_def}
\end{equation}
\begin{equation}
    \hat{L}_{(x,y)} \big{[}B_n(k_0\rho)e^{-jn\phi}\big{]} = B_{n-1}(k_0\rho)e^{-j(n-1)\phi}.
\label{eq_App_Lowering_operation}
\end{equation}

To see why (\ref{eq_App_Raising_operation}) is true, its partial derivatives need to be carried out explicitly. By utilizing the following relationships,
\begin{equation}
    \frac{\partial \rho}{\partial x} = \frac{x}{\sqrt{x^2+y^2}} = \cos\phi, \quad
    \frac{\partial \rho}{\partial y} = \frac{y}{\sqrt{x^2+y^2}} = \sin\phi,
\end{equation}
\begin{equation}
    \frac{\partial \phi}{\partial x} = -\frac{y}{x^2+y^2} = -\frac{1}{\rho}\sin\phi,
\end{equation}
\begin{equation}
    \frac{\partial \phi}{\partial y} = +\frac{x}{x^2+y^2} = +\frac{1}{\rho}\cos\phi,
\end{equation}
\noindent{it can be shown that,}
\begin{equation}
\begin{split}
    &\hat{R}_{(x,y)} \big{[}B_n(k_0\rho)e^{-jn\phi}\big{]} \\
    &= -B_n'(k_0\rho)e^{-j(n+1)\phi} + \frac{n}{k_0\rho}B_n(k_0\rho)e^{-j(n+1)\phi}.
\end{split}
\end{equation}
\noindent{The right hand side of (\ref{eq_App_Raising_operation}) follows directly from this equation once the
following Bessel function recurrence relation \cite{Chew_book, Harrington_book} is applied:}
\begin{equation}
    B_n'(k_0\rho) = -B_{n+1}(k_0\rho) +\frac{n}{k_0\rho}B_n(k_0\rho).
\label{eq_App_Recurrence_1}
\end{equation}
\noindent{Likewise, (\ref{eq_App_Lowering_operation}) can also be demonstrated by incorporating another Bessel function
recurrence relation:}
\begin{equation}
    B_n'(k_0\rho) = +B_{n-1}(k_0\rho) -\frac{n}{k_0\rho}B_n(k_0\rho).
\label{eq_App_Recurrence_2}
\end{equation}

Now that the raising and lowering operators are defined, the proof of the addition theorem for outgoing cylindrical waves (\ref{eq_2B_Hankel2_addition}) is straightforward. We start from the well-known Hankel function addition theorem for the fundamental mode ($m=0$) \cite{Chew_book, Martin_book, Harrington_book},
\begin{equation}
\begin{split}
    &H_0^{(2)}(k_0\rho'') = H_0^{(2)}(k_0|\bar{\rho}-\bar{\rho}'|) \\
    &=
    \begin{cases}
      \sum_n J_{n}(k_0\rho')H_n^{(2)}(k_0\rho)e^{-jn\phi}e^{+jn\phi'}& \text{if}\ \rho \geq \rho' \\
      \sum_n H_{n}^{(2)}(k_0\rho')J_n(k_0\rho)e^{-jn\phi}e^{+jn\phi'}& \text{if}\ \rho < \rho'
    \end{cases}
\end{split}
\label{eq_App_Hankel2_addition}
\end{equation}
\noindent{Next, let us apply the following raising operator $m$ times to both sides of (\ref{eq_App_Hankel2_addition}):}
\begin{equation}
    \hat{R}_{(x'',y'')} =-\frac{1}{k}\Big{(} \frac{\partial}{\partial x''} - j\frac{\partial}{\partial y''} \Big{)},
\end{equation}
\noindent{where $x''$ and $y''$ represent the rectangular coordinates of $(\rho'',\phi'')$, the point of interest
in Fig. \ref{fig_add} relative to the displaced feed.
By definition $x'=\rho'\cos\phi'$ and $y'=\rho'\sin\phi'$ denote the rectangular coordinates of the displaced feed.
Therefore,}
\begin{equation}
\begin{alignedat}{3}
    x'' &=\rho''\cos\phi'' &&= x - x' &&= \rho\cos\phi - \rho'\cos\phi' \\
    y'' &=\rho''\sin\phi'' &&= y - y' &&= \rho\sin\phi - \rho'\sin\phi'.
\end{alignedat}
\label{eq_App_xxpp_relation}
\end{equation}
\noindent{When $\rho\geq\rho'$, (\ref{eq_App_Hankel2_addition}) becomes,}
\begin{equation}
\begin{split}
    &\big{[}\hat{R}_{(x'',y'')}\big{]}^m \Big{\{} H_0^{(2)}(k_0\rho'') \Big{\}} \\
    &= \big{[}\hat{R}_{(x'',y'')}\big{]}^m \Big{\{} \sum_n J_{n}(k_0\rho')H_n^{(2)}(k_0\rho)e^{-jn\phi}e^{+jn\phi'} \Big{\}}.
\label{eq_App_Hankel2_addition_mod1}
\end{split}
\end{equation}
\noindent{Since the location of the displaced feed is given and fixed, $x'$ and $y'$ are constants throughout the analysis.
It is now simple using (\ref{eq_App_xxpp_relation}) to prove that $\hat{R}_{(x'',y'')} = \hat{R}_{(x,y)}$:}
\begin{equation*}
\begin{split}
    \hat{R}_{(x'',y'')} &= -\frac{1}{k}\Big{(} \frac{\partial}{\partial x''} - j\frac{\partial}{\partial y''} \Big{)}
    \\
    &= -\frac{1}{k} \bigg[
    \Big{(} \frac{\partial x}{\partial x''} \Big{)} \frac{\partial}{\partial x} - j
    \Big{(} \frac{\partial y}{\partial y''} \Big{)}
    \frac{\partial}{\partial y}
    \bigg]
    \\
    &= -\frac{1}{k} \bigg[
    (1) \frac{\partial}{\partial x} - j
    (1) \frac{\partial}{\partial y}
    \bigg] \quad\quad\quad\quad = \hat{R}_{(x,y)}.
\end{split}
\end{equation*}

\noindent{Therefore, the right-hand-side of (\ref{eq_App_Hankel2_addition_mod1}) can be rewritten as:}
\begin{equation}
\begin{split}
    &\big{[}\hat{R}_{(x'',y'')}\big{]}^m \Big{\{} H_0^{(2)}(k_0\rho'') \Big{\}} \\
    &= \sum_n J_{n}(k_0\rho')e^{+jn\phi'} \big{[}\hat{R}_{(x,y)}\big{]}^m \Big{\{} H_n^{(2)}(k_0\rho)e^{-jn\phi} \Big{\}}.
\label{eq_App_Hankel2_addition_mod2}
\end{split}
\end{equation}
\noindent{This relation indicates that the raising operators directly yield,}
\begin{equation}
\begin{split}
    &H_m^{(2)}(k_0\rho'') e^{-jm\phi''} \\
    &= \sum_n J_{n}(k_0\rho')e^{+jn\phi'} H_{n+m}^{(2)}(k_0\rho)e^{-j(n+m)\phi}.
\label{eq_App_Hankel2_addition_mod3}
\end{split}
\end{equation}
\noindent{Finally, by changing the summation variable $n$ on the right-hand-side to $n-m$, the first part ($\rho\geq\rho'$) of (\ref{eq_2B_Hankel2_addition}) is proven,}
\begin{equation}
\begin{split}
    &H_m^{(2)}(k_0\rho'') e^{-jm\phi''} \\
    &= \sum_n J_{n-m}(k_0\rho')e^{+j(n-m)\phi'} H_{n}^{(2)}(k_0\rho)e^{-jn\phi}.
\label{eq_App_Hankel2_addition_mod4}
\end{split}
\end{equation}
\noindent{The case of $\rho<\rho'$ can be also proven by adopting the same procedure. This finishes the proof of the
addition theorem for outgoing waves (\ref{eq_2B_Hankel2_addition}).}

The addition theorem for incoming waves (\ref{eq_2B_Hankel1_addition}) can be derived directly by complex conjugating (\ref{eq_2B_Hankel2_addition}), i.e.,
\begin{equation}
\begin{split}
    &H_m^{(1)}(k_0\rho'')e^{+jm\phi''} \\
    &=
    \begin{cases}
      \sum_n J_{n-m}(k_0\rho')H_n^{(1)}(k_0\rho)e^{+jn\phi}e^{-j(n-m)\phi'}& \text{if}\ \rho \geq \rho' \\
      \sum_n H_{n-m}^{(1)}(k_0\rho')J_n(k_0\rho)e^{+jn\phi}e^{-j(n-m)\phi'}& \text{if}\ \rho < \rho'
    \end{cases}
\end{split}
\label{eq_App_Hankel1_addition_mod1}
\end{equation}
\noindent{Next, by changing the variables $m$ and $n$ in (\ref{eq_App_Hankel1_addition_mod1}) to $-m$ and $-n$, respectively, leads to,}
\begin{equation}
\begin{split}
    &H_{-m}^{(1)}(k_0\rho'')e^{-jm\phi''} \\
    &=
    \begin{cases}
      \sum_n J_{m-n}(k_0\rho')H_{-n}^{(1)}(k_0\rho)e^{-jn\phi}e^{+j(n-m)\phi'}& \text{if}\ \rho \geq \rho' \\
      \sum_n H_{m-n}^{(1)}(k_0\rho')J_{-n}(k_0\rho)e^{-jn\phi}e^{+j(n-m)\phi'}& \text{if}\ \rho < \rho'
    \end{cases}
\end{split}
\label{eq_App_Hankel1_addition_mod2}
\end{equation}
\noindent{Then, by utilizing the following property of $B_n$ \cite{Harrington_book}:}
\begin{equation}
    B_{-n}(k_0\rho) = (-1)^n B_n(k_0\rho),
\end{equation}
\noindent{(\ref{eq_App_Hankel1_addition_mod2}) can be rewritten as:}
\begin{equation}
\begin{split}
    &(-1)^m H_m^{(1)}(k_0\rho'')e^{-jm\phi''} \\
    &=
    \begin{cases}
      \sum_n
      (-1)^{m-n} J_{n-m}(k_0\rho')\\
      \hspace{0.6cm} (-1)^n H_n^{(1)}(k_0\rho)e^{-jn\phi}e^{+j(n-m)\phi'}& \text{if}\ \rho \geq \rho' \\
      \sum_n
      (-1)^{m-n} H_{n-m}^{(1)}(k_0\rho')\\
      \hspace{0.6cm} (-1)^n J_n(k_0\rho)e^{-jn\phi}e^{+j(n-m)\phi'}& \text{if}\ \rho < \rho'
    \end{cases}
\end{split}
\label{eq_App_Hankel1_addition_mod3}
\end{equation}
\noindent{Finally, (\ref{eq_2B_Hankel1_addition}) is proven by cancelling out the $(-1)^m$ on both sides of the equation.} 


\end{document}